\documentclass[prl,aps,twocolumn,superscriptaddress,showpacs,footinbib,longbibliography]{revtex4-1}
\usepackage[colorlinks=true,linkcolor=black, citecolor=blue, urlcolor=blue, 
unicode=true,breaklinks]{hyperref}
\usepackage{graphicx}
\usepackage{epstopdf}
\usepackage{graphicx}
\usepackage{amsmath}
\usepackage{dsfont} 
\usepackage{pgfplots}
\usepackage{physics}
\usepackage{array}
\usepackage{svg}

\usepackage{amssymb}
\usepackage{commath}

\usepackage{verbatim}

\begin{document}

\title{Zeeman split Kramers doublets in spin-supersolid candidate Na$_{2}$BaCo(PO$_{4}$)$_{2}$}
\author{T. I. Popescu}
\affiliation{School of Physics and Astronomy, University of Edinburgh, EH9 3JZ, United Kingdom}
\author{N. Gora}
\affiliation{School of Physics and Astronomy, University of Edinburgh, EH9 3JZ, United Kingdom}
\author{F. Demmel}
\affiliation{ISIS Pulsed Neutron and Muon Source, STFC Rutherford Appleton Laboratory, Harwell Campus, Didcot, Oxon OX11 0QX, United Kingdom}
\author{Z. Xu}
\affiliation{NIST Center for Neutron Research, 100 Bureau Drive, Gaithersburg, Maryland 20899, USA}
\author{R. Zhong}
\affiliation{Department of Chemistry, Princeton University, Princeton NJ 08544, USA}
\author{T. J. Williams}
\affiliation{ISIS Pulsed Neutron and Muon Source, STFC Rutherford Appleton Laboratory, Harwell Campus, Didcot, Oxon OX11 0QX, United Kingdom}
\author{R. J. Cava}
\affiliation{Department of Chemistry, Princeton University, Princeton NJ 08544, USA}
\author{G. Xu}
\affiliation{NIST Center for Neutron Research, 100 Bureau Drive, Gaithersburg, Maryland 20899, USA}
\author{C. Stock}
\affiliation{School of Physics and Astronomy, University of Edinburgh, EH9 3JZ, United Kingdom}

\date{\today}

\begin{abstract}

Na$_{2}$BaCo(PO$_{4}$)$_{2}$ is a triangular antiferromagnet that displays highly efficient adiabatic demagnetization cooling (Ref. \onlinecite{Xiang24:625}) near a quantum critical point at $\mu_{0}H_{c}\sim 1.6$ T, separating a low-field magnetically disordered from a high-field fully polarized ferromagnetic phase.   We apply high resolution backscattering neutron spectroscopy in an applied field to study the magnetic excitations near $\mu_{0}H_{c}$.  At large fields we observe ferromagnetic fluctuations that gradually transition to being overdamped in energy below $\mu_{0}H_{c}$ where the magnetism is spatially disordered.  We parameterize the excitations in the high field polarized phase in terms of coupled Zeeman split Kramers doublets originating from the presence of spin-orbit coupling.  On reducing the field, the splitting between the Kramers doublets is reduced and if done adiabatically, provides a mechanism for reducing temperature.  On lowering the applied field through the $\mu_{0}H_{c}$ the excitations characterize a textured phase that we suggest is inefficient for cooling.  Low temperature disordered frustrated magnets built on Kramers doublets with nearby quantum critical points provide a route for efficient magnetocalorics.

\end{abstract}

\pacs{}
\maketitle

Low-dimensional magnets with frustrating interactions provide a platform for novel quantum phases and properties~\cite{Collins97:75}.  While the classical two-dimensional triangular antiferromagnetic displays 120$^{\circ}$ magnetic order, other exotic phases have been predicted to exist depending on the magnetic spin including ``hidden" nematic quadrupolar orders~\cite{Tsunetsugu06:75,Stoudenmire09:79}, disordered spin-liquid phases~\cite{Anderson73:8}, and recently proposed spin-supersolids~\cite{Wessel05:95,Melko05:95,Heidarian05:95,Boninsegni05:95,Heidarian10:104,Wang09:102,Jiang09:79} analogous to the supersolid phase sought after in solid Helium~\cite{Kim04:427}.  Here we investigate the magnetic excitations in a candidate spin-supersolid~\cite{Gao22:89} and characterize the low temperature magnetic excitations as a function of magnetic field and discuss the response in terms of recently reported efficient adiabatic demagnetization cooling.

\begin{figure} [ht]
	\includegraphics[width=85mm,trim=0cm 0cm 0cm 0cm,clip=true]{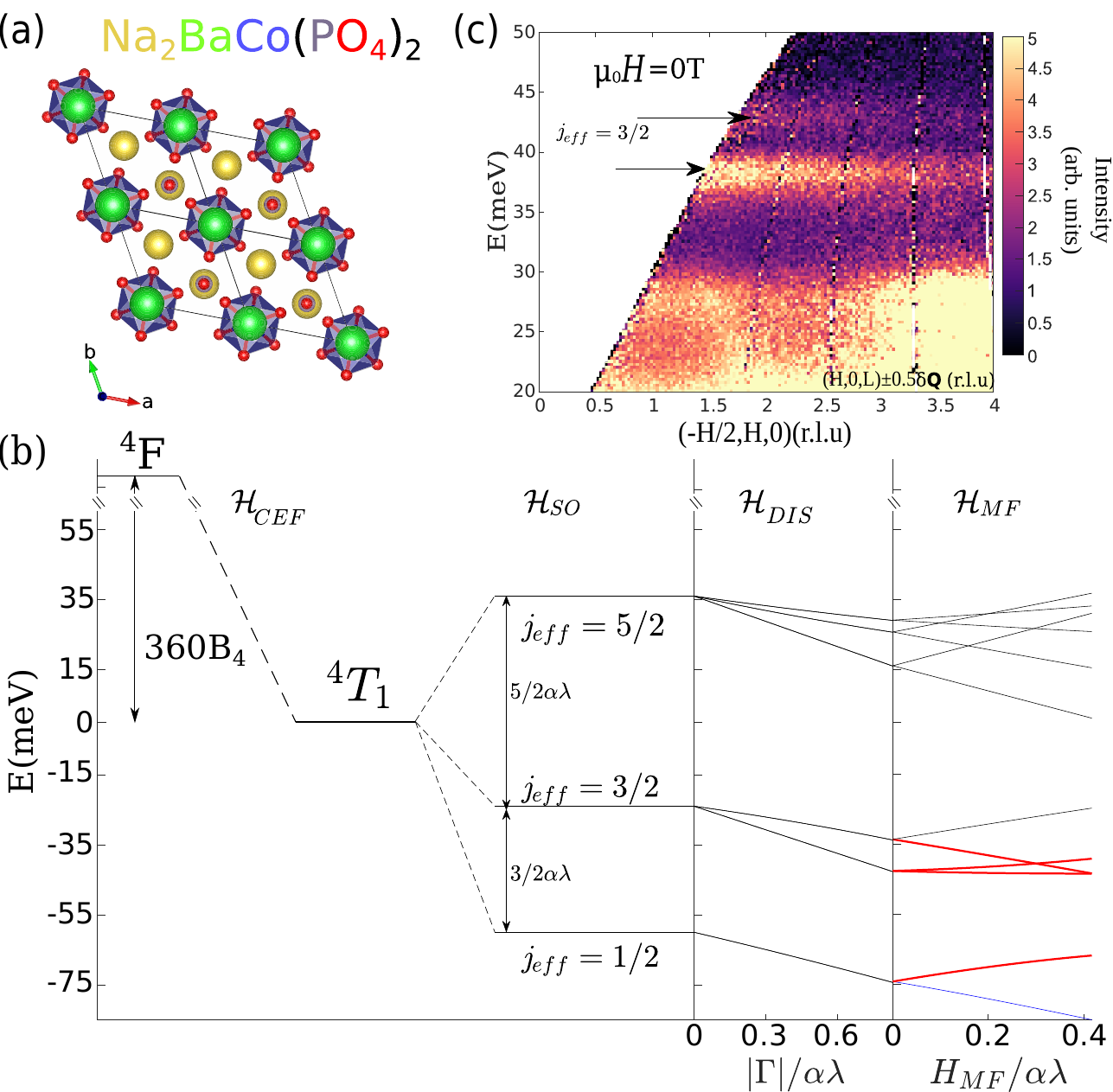}
	\caption{$(a)$ Structure of Na$_{2}$BaCo(PO$_{4}$)$_{2}$ (NBCP) with octahedral CoO$_6$. $(b)$ Co$^{2+}$ single-ion energies under spin-orbit, distortion and molecular field with an overall applied octahedral crystalline electric field. The dipole active quantum levels with finite neutron cross section from the ground state (blue) are highlighted in red. $(c)$ Constant momentum slice (MAPS) at T=10 K of the first spin-orbit level $j_{eff}=3/2$. We note that given the direction of the integration, the spectral intensities found at zero on the $x$-axis describe a sum over (0,0,$\pm$0.5) rather than data strictly at the $|\vec{Q}|$=0 origin.}
	\label{figure_1}
\end{figure}

Antiferromagnetic triangular magnets can often be fully polarized with relatively small applied magnetic fields through a metamagnetic transition.~\cite{Stryjewski77:26}  In terms of applications, the critical field separating a low-field disordered spin phase to a fully polarized state has been observed to host efficient magnetocaloric properties making such systems candidates for adiabatic demagnetization.~\cite{Xiang24:625,Zhang24:8}  In particular, Na$_{2}$BaCo(PO$_{4}$)$_{2}$ (NBCP) has been identified to display giant magnetocaloric properties.~\cite{Xiang24:625}  NBCP consists of two-dimensional planes of coupled Co$^{2+}$ ions (Fig. \ref{figure_1} $a$) in triangular arrangement resulting from a trigonal $P\overline{3}m1$ structure with lattice parameters $a=b=5.3185$ \AA\ and $c=$7.0081 \AA.  Despite a Curie-Weiss constant of $|\Theta|_{CW}\sim$ 2 K, magnetic order only appears at T$_{N} \sim$ 150 mK with weak (reduced moment of $\mu\sim 0.5$ $\mu_{B}$) incommensurate magnetic order.~\cite{Sheng22:119,Lee21:103}  On application of a magnetic field, this incommensurate magnetic order is replaced by an up-up-down $|\hspace{-4pt}\uparrow\uparrow\downarrow\rangle$ order with propagation wavevector $\vec{q}_{0}=({1\over3},{1\over3})$ and has been proposed to host a spin-supersolid phase~\cite{Zhong19:116,Li20:11} supported by theoretical analysis of low-field neutron spectroscopy.~\cite{Gao24:110}  At large magnetic fields greater than $\mu_{0} H_{c}\sim$1.6 T, a fully polarized $|\hspace{-2pt}\uparrow \uparrow \uparrow \rangle$ magnetic state is present.

The critical magnetic field separating $|\hspace{-2pt}\uparrow\uparrow\downarrow\rangle$ and $|\hspace{-2pt}\uparrow \uparrow \uparrow \rangle$ phases has been found to display very efficient adiabatic demagnetization cooling.~\cite{Xiang24:625}  In particular, reducing the field from large fields towards the critical field of $\mu_{0} H_{c}\sim$1.6 T has been found to reduce the temperature of the sample from a $T_{0}$=2 K to $T$ $<$ 0.1 K.  However, for lower fields within the low field $|\hspace{-2pt}\uparrow\uparrow\downarrow\rangle$ phase warming is observed and eventual recovery of low temperatures found as the applied magnetic field is reduced to zero.

\begin{figure*} [t]
	\includegraphics[width=170mm]{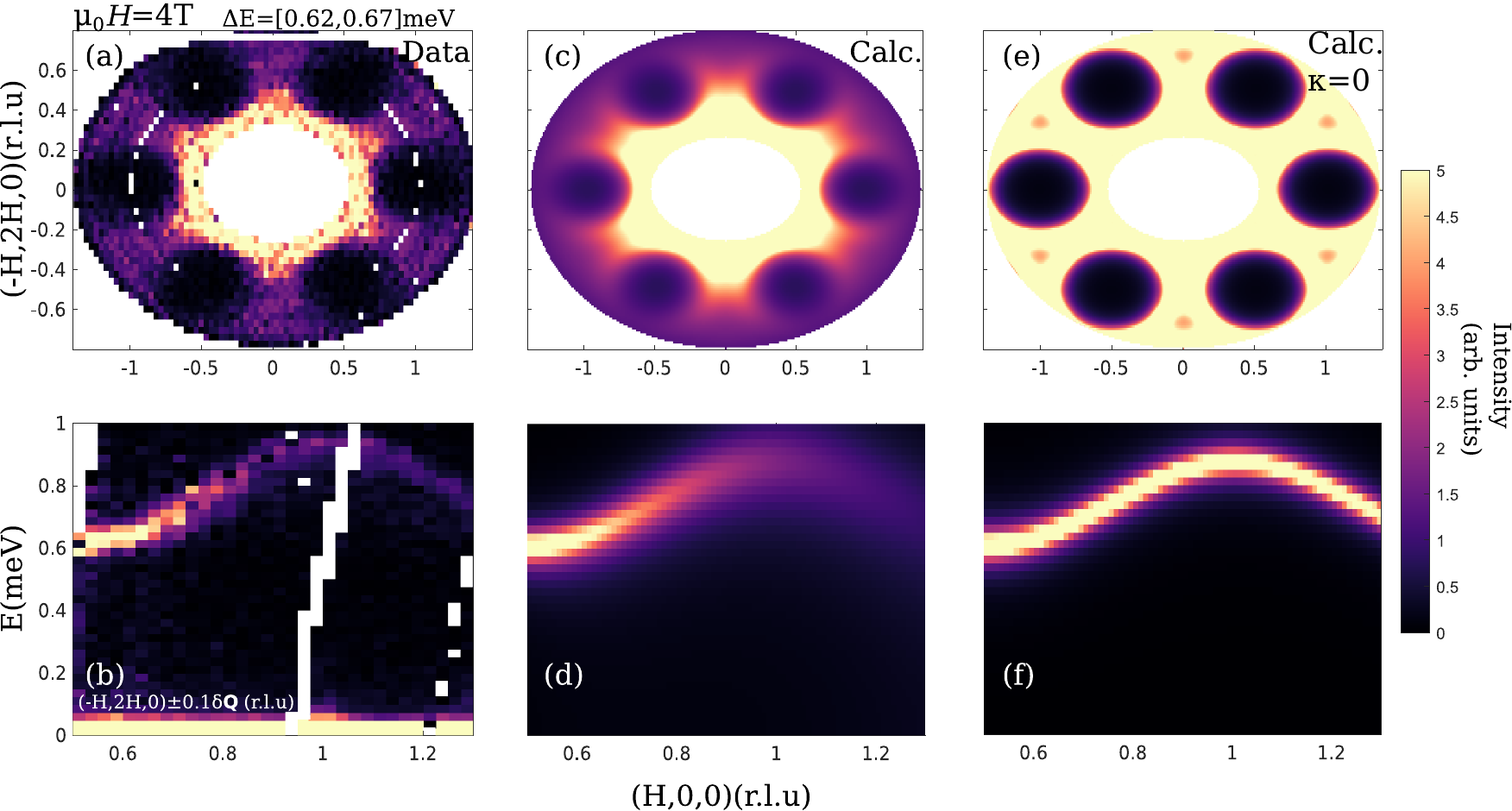}
	\caption{Constant energy slices (OSIRIS) taken at 4~T from (a) the measured data between $\Delta E=[0.62,0.67]$ meV, (c) RPA calculations with the inclusion of the term $\kappa$ taken at E=0.645 meV and (e) RPA theory excluding $\kappa$. Constant momentum slices are also presented with (b) data cuts along the (-H$\pm$0.1, 2H$\pm$0.1, 0) direction. Additional RPA theoretical dispersion plots taken at H=0 along the (-H,2H,0) direction (d) including the parameter $\kappa$ and (f) excluding it.}
	\label{figure_2}
\end{figure*}

Cooling through adiabatic demagnetization involves splitting a ground state degeneracy with an applied magnetic field preferentially depopulating the energetically costly excited states followed by lowering the field adiabatically (keeping relative populations constant). This can be understood through the Boltzmann distribution, $n(E)\propto e^{-\Delta E/k_b T}$, where demagnetization reduces $\Delta E$ resulting in a lowering of the temperature of the system to maintain a constant particle number at each energy level $n(E)$.  In this context, the foundation relevant for NBCP are the magnetic Co$^{2+}$ ions coordinated in an octahedral environment.  Given the multi level quantum nature of demagnetization cooling, we first discuss the single-ion physics of isolated Co$^{2+}$ ions in applied magnetic fields and then discuss experiments and data parameterization.  

The magnetic Hamiltonian for Co$^{2+}$ ions in NBCP can be divided into single-ion (SI) and interacting parts,

\begin{equation}
	\mathcal{H} = \mathcal{H}_{\text{SI}} + \sum_{ij} J(ij) \boldsymbol{S}(i) \cdot \boldsymbol{S}(j). \nonumber
\end{equation}

\noindent We first discuss the single-ion Hamiltonian $\mathcal{H}_\text{SI}$. The hierarchy and effects of different terms in the single-ion Hamiltonian for Co$^{2+}$ (7 $d$-electrons) are illustrated in Fig. \ref{figure_1} $(b)$.~\cite{Sakurai68:167}  The largest single-ion energy scale is the crystalline electric field $\mathcal{H}_{CEF}$ from an octahedral environment of oxygen that stabilizes a ground state spin-orbital triplet ($^{4}T_{1}$, Fig. \ref{figure_1}) with an effective angular momentum of $l_{eff}=1$ and spin $S={3\over2}$.~\cite{Cowley13:88}  Given the presence of an orbital degeneracy, the next important term is spin orbit coupling $\mathcal{H}_{SO}=\alpha \lambda \vec{l}\cdot\vec{S}$ ($\alpha=-{3\over2}$ is a projection factor~\cite{Cowley13:88,Sarte18:98}) which splits the spin-orbital states into three levels with total angular momentum $j_{eff}={5\over 2}$, ${3\over2}$, and a ground state ${1\over2}$. The local octahedra surrounding the Co$^{2+}$ site is distorted in NBCP and we parameterize this deviation of the crystalline electric field by a tetragonal distortion $\mathcal{H}_{dis}=\Gamma (l_{z}^{2}-{2\over3})$.  The ground state degeneracy of the Kramers doublet is broken by a time reversal symmetry breaking field $\mathcal{H}_{MF}=h_{mf}S_{z}$ either by an applied magnetic field or a local molecular field from neighboring statically ordered spins.  

In the following we parameterize the magnetic excitations extracted from neutron spectroscopy based on the single-ion eigenstates of $\mathcal{H}_{SI}$.  We apply Green's response functions as the neutron scattering cross section is proportional to the imaginary part of the response function.  This follows previous work applying this methodology to compounds with spin-orbit coupling~\cite{Sarte19:100,Sarte20:102,Lane22:106,Lane23:5,Lane21:104}, spin-only~\cite{Chan23:107}, and rare-earth~\cite{Brener24:110} compounds and is outlined in the Supplementary Information (SI).  The calculations are based on mean-field theory applying the random phase approximation (RPA)~\cite{Buyers75:11}

\begin{equation}
	{G}({\bf{Q}},\omega)={{g}}(\omega) \left[ {{\mathds{1}}}-{\mathcal{J}}({\bf{Q}}) {{g}}(\omega) \right]^{-1}.\nonumber
\end{equation}

\noindent with single-site susceptibility $g(\omega)$ derived from the eigenstates of $\mathcal{H}_\text{SI}$.  The Fourier transform of the exchange constants is denoted as $\mathcal{J}({\bf{q}})$.

Neutron spectroscopy was performed on OSIRIS~\cite{Telling05:7} (ISIS, UK) with a coaligned NBCP sample mounted in a dilution fridge with (HK0) Bragg reflections in the horizontal plane.  The elastic energy resolution with $E_{f}$=1.84 meV is 0.025 meV (full-width)~\cite{Demmel14:767}.  While the mixing chamber was held at T=50 mK,  no evidence for static magnetism was observed and this taken with previous calibrations performed on heavy fermions~\cite{Stock12:109,Stock12:109_2} in similar configurations leads us to the conclusion the sample temperature was greater than $\sim$ 100 mK.  Vertical resolution defined by the detector geometry is estimated to be $\pm$ 0.2 r.l.u along $c^{*}$.  Higher energy data characterizing spin-orbit transitions were taken on the MAPS spectrometers.  Further experimental details are given in the SI.

We now discuss our MAPS data (Fig. \ref{figure_1} $c$, T=10 K) where the only terms that are expected to define $\mathcal{H}_{SI}$ are the distortion $\Gamma$ and spin-orbit coupling $\lambda$.  The dominant cross section in Fig. \ref{figure_1} (below $\sim$ 30 meV) are phonons as the scattering increases with momentum transfer.  Two distinct dispersionless bands near $\sim$ 40 meV match the expected splitting of energy levels in Fig. \ref{figure_1} $(b)$ for excitations from the ground state $j_{eff}={1\over 2} \rightarrow j_{eff}={3\over 2}$ manifold. As discussed in the SI, the intensity from these excitations decays with increasing momentum transfer and is described by the isotropic Co$^{2+}$ form factor~\cite{Kernavanois04:350,Kernavanois03:15}.   The energy position of these two modes fixes $\lambda$=-18 $\pm$ 2 meV and $\Gamma$=-10 $\pm$ 2 meV in $\mathcal{H}_{SI}$.

We now discuss the low-energy excitations within the $j_{eff}={1\over 2}$ ground state and our parameterization of the magnetic excitations in the low-temperature fully spin-polarized $|\hspace{-4pt}\uparrow \uparrow \uparrow \rangle$ phase.  Figs. \ref{figure_2} $(a)$ and $(b)$ illustrate constant energy and momentum slices through the magnetic excitations which disperse over the energy range of $\sim$0.6-0.95 meV. We describe these in terms of single-ion Kramers doublets in  Figs. \ref{figure_2} $(e,f)$ coupled with a nearest neighbor exchange coupling of $J$=0.0095 meV via RPA.  The calculated neutron response includes an isotropic Co$^{2+}$ form factor found to describe the momentum dependent intensity of the single-ion excitations in Fig. \ref{figure_1} $(c)$.  The derived exchange constant can be used to calculate a Curie-Weiss constant (assuming $j_{eff}={1\over 2}$) of $|\Theta_{CW}|$=2 K, which is in agreement with experiment. However, while the RPA (Figs. \ref{figure_2} $e,f$) describes the magnetic dispersion and momentum dependence, it fails to describe the momentum dependent intensity (Figs. \ref{figure_2} $a,b$), particularly at large momentum transfers near the zone boundary where the intensity in the data decreases faster with momentum transfer than would be expected from the isotropic Co$^{2+}$ magnetic form factor.  

\begin{figure} [t]
	\includegraphics[width=85mm]{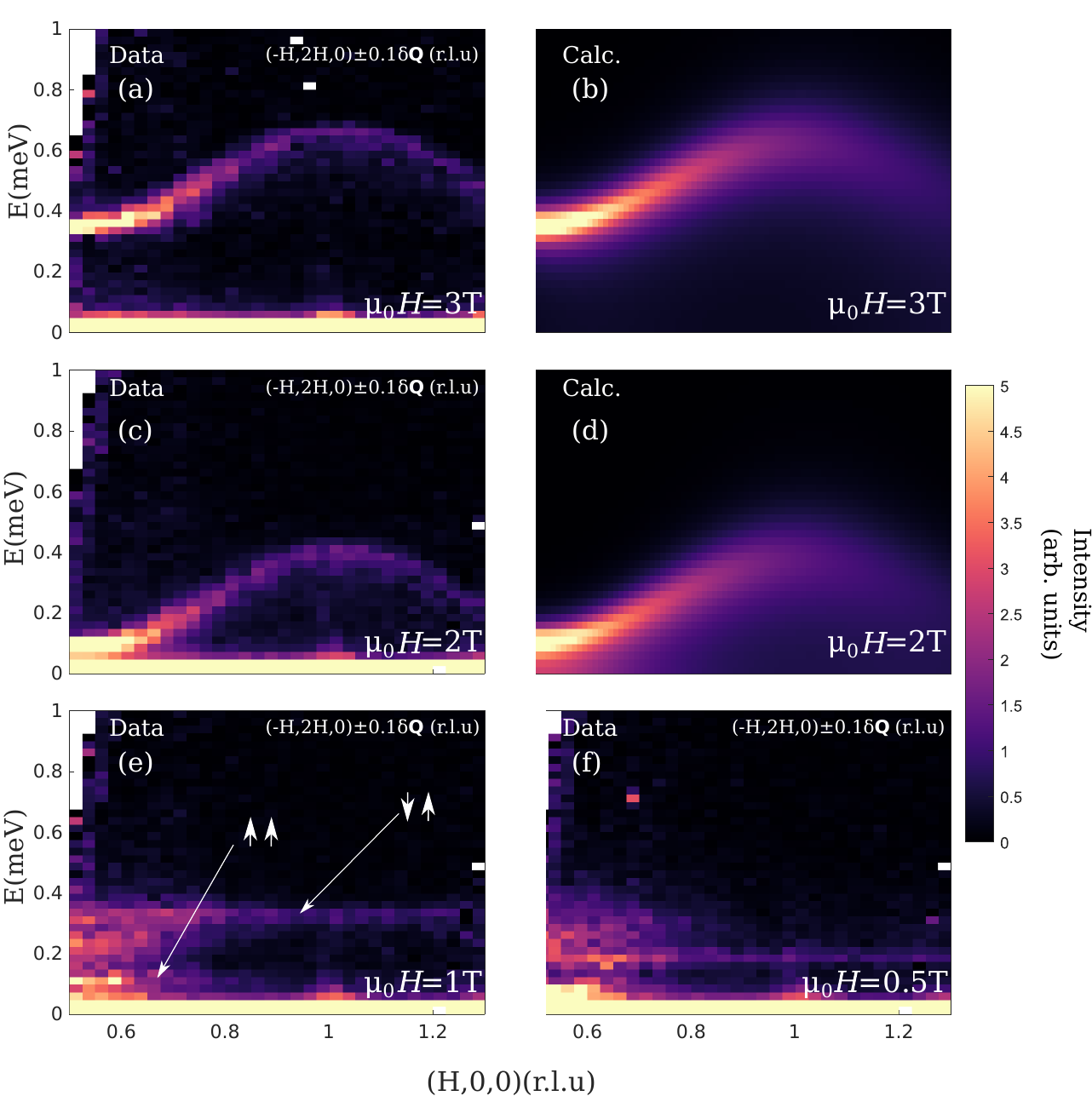}
	\caption{Constant momentum slices (OSIRIS) along (-H$\pm$0.1, 2H$\pm$0.1, 0) for $(a)$ $\mu_0\mathit{H}$ = 3.0 T, (c) 2.0 T, (e) 1.0 T, (f) 0.5 T. RPA calculations for (b) $\mu_0\mathit{H}$=3.0 T and (d) 2.0 T.}
	\label{figure_3}
\end{figure}

To account for this decrease in intensity, we follow Ref. \onlinecite{Zhitomirsky13:85,Masuda06:96} and consider a distribution of response functions centered around the Green's response function predicted with the RPA calculation outlined above.  To account for this and motivated by the ``imaginary-part Self Consistent Born Approximation (iSCBA)" we include an additional parameter $\kappa$

\begin{equation}
\mathcal{G}^{-1}(\mathbf{Q},\omega)=G^{-1}(\mathbf{Q},\omega)+i\kappa.
\end{equation}

\noindent The extra term $\kappa$ results in a broadening of the excitations.  Following works on liquids~\cite{Stanley:book} we assign a momentum dependence to this term that varies to leading order $\kappa=\xi |\mathbf{Q}|^{2}$, preserving even symmetry in momentum transfer.  Inclusion of this heuristic parameter $\xi$ that characterizes the distribution of Green's response functions is illustrated in Fig. \ref{figure_2} $(c,d)$ and accounts for the decrease of intensity with increasing momentum transfer. We discuss the field dependence of this parameter below.  This additional term can be motivated by a mode-coupling analysis~\cite{Wehner75:36} which may result from the coupling of single-magnon excitations to another degree of freedom such as a multi-magnon continuum.  While the single and multi magnon dispersion relations do not kinematically intersect, the two process are expected to become close energetically near the zone boundary thereby potentially enhancing coupling. 

The field dependence of the dispersive excitations is illustrated in Fig. \ref{figure_3}.  Decreasing applied magnetic fields of $\mu_{0}H$=3, 2 T is shown in Figs. \ref{figure_3} $(a,c)$ where the excitation energetic gap decreases. This is replicated in our response theory in Figs. \ref{figure_3} $(b,d)$ and can be understood in terms of a decrease in splitting of the ground state $j_{eff}={1\over 2}$ Kramers doublet illustrated in the single-ion response in Fig. \ref{figure_1} $(b)$.  This originates from a combination of a decrease in the applied magnetic field and decrease of local magnetic order contributing to the molecular field term in the Fig. \ref{figure_1} $(b)$.  This reduced splitting of the ground state Kramers doublet, if performed adiabatically, keeping the relative population of the differing quantum levels fixed, will lead to cooling of the sample.

\begin{figure}[t]
	\includegraphics[ width=1\linewidth]{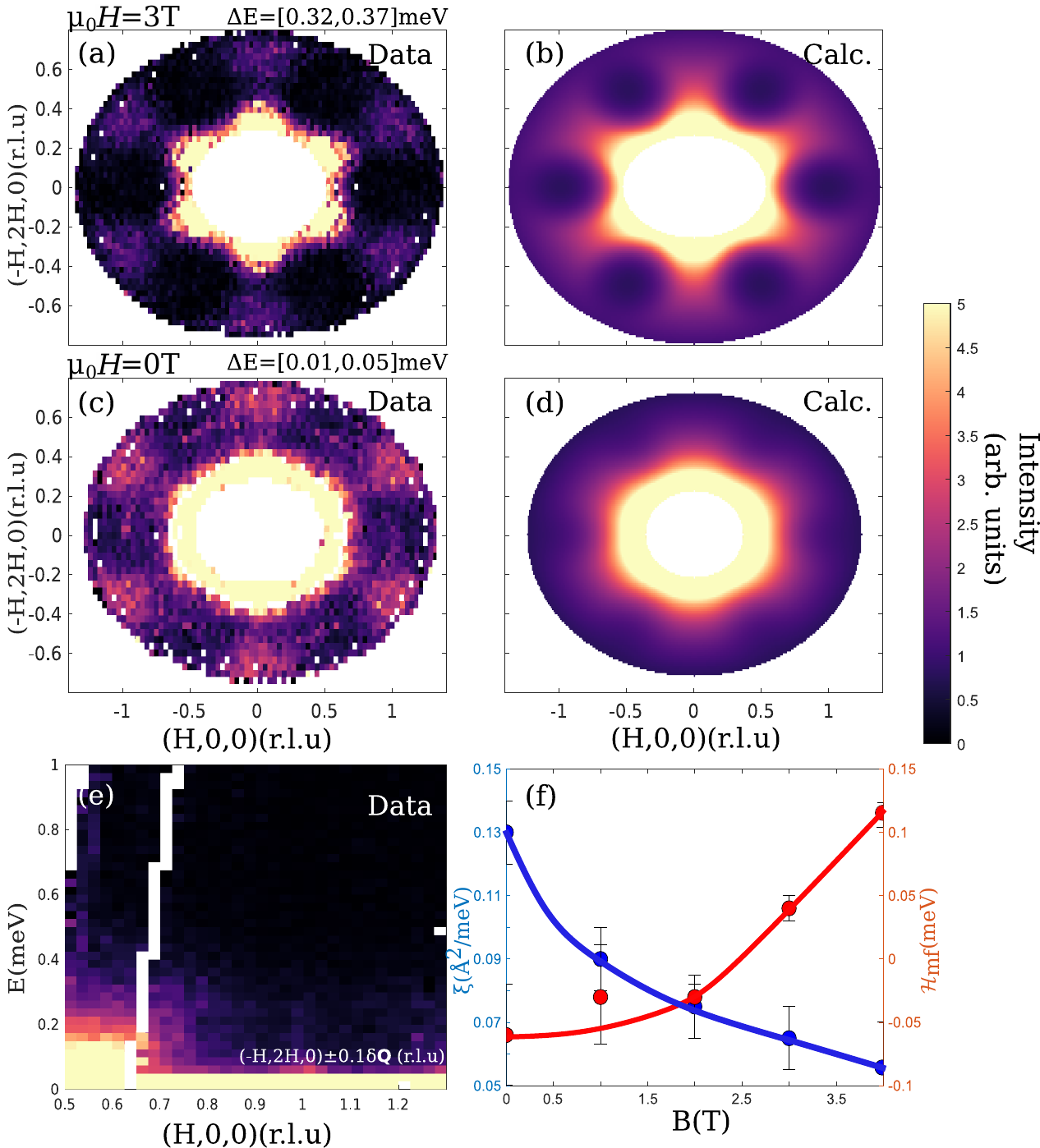}
	\caption{Folded constant-$E$ slices (OSIRIS) at $(a)$ E=[0.32,0.37] meV (3 T) and $(c)$ [0.01,0.015] meV (0 T). RPA calculations at $(b)$ E=0.325 meV (3 T) and $(d)$ 0.085 meV (0 T). $(e)$ Constant-$\vec{Q}$ slice (OSIRIS) at $\mu_{0}H$= 0 T. $(f)$ Obtained parameters as a function of $\mu_{0}H$ with guiding lines.}
	\label{figure_4}
\end{figure}

Below 2 T and illustrated in Figs. \ref{figure_3} $(e,f)$, the neutron response changes dramatically from leaving the fully polarized $|\hspace{-3pt}\uparrow \uparrow \uparrow \rangle$ phase to the intermediate $|\hspace{-3pt}\uparrow \uparrow \downarrow \rangle$ phase.  Figs. \ref{figure_3} $(e,f)$ are complex consisting of correlated magnetic scattering near H $\sim$ 0.5 and also two field dependent flat modes that soften as the field is lowered from 1 to 0.5 T. It is possible to attribute the two visible flat modes to pair spectra of nearest neighbor Co$^{2+}$ ions. The higher energy flat mode in Figs. \ref{figure_3} $(e,f)$ can be assigned to a pair excitation of anti-aligned spins, with one of the spins experiencing a strong molecular field from its six nearest neighbors being aligned with the field. Meanwhile, the lower energy mode originates from a pair excitation of aligned spins, this lower energy originating from the spins not experiencing a molecular field. Both situations are expected to exist in the textured $|\hspace{-3pt}\uparrow \uparrow \downarrow \rangle$ phase. These excitations are indicative of a complex or textured order which may be expected given Ising-like $|\hspace{-2pt}\uparrow \downarrow \rangle$ on a triangular motif is likely disordered. The complex and multi level excitation spectrum is indicative of a large entropic ground state which is disordered.  It is in this $|\hspace{-2pt}\uparrow \uparrow \downarrow \rangle$ phase that NBCP lacks cooling efficiency.

In Fig. \ref{figure_4} we compare constant energy slices in this intermediate phase (Figs. \ref{figure_4} $a,c$) to our parameterization outlined in Figs. \ref{figure_4} $(b,d)$.  On lowering the field and on comparing with Fig. \ref{figure_2} the momentum dependence in Figs. \ref{figure_4} $(a,c)$ became broader, indicative of a shortening of dynamic spatial correlations in the $|\hspace{-4pt}\uparrow \uparrow \downarrow \rangle$ and the $\mu_{0}H$=0 T phases.  The loss of spatial correlations is particularly evident in Fig. \ref{figure_4} $(c)$ at small momentum transfers where the scattering nearly has lost the sixfold symmetry seen at higher applied fields and forms nearly a ring in momentum.  Simultaneously, the temporal correlations (Fig. \ref{figure_4} $e$) become extended at zero applied field.  This relaxational-like scattering at $\mu_{0}H$=0 T is in contrast to the dispersive excitations discussed above in the polarized phase for fields in excess of $\mu_{0}H_{c}$.  

The momentum broadening is reproduced in our calculations in Figs. \ref{figure_4} $(b,d)$ by increasing the parameter $\kappa$ which corresponds to a broader distribution of Green's functions discussed above. In Fig. \ref{figure_4} $(f)$ we plot the parameters $\xi$ that characterizes the distribution of response functions and also the molecular field we have used as a function of applied magnetic field.  There is a trade off between the two parameters with a nearly fully saturated molecular field found to reproduce the magnetic response at $\mu_{0}H$=4 T and minimal $\xi$. Indeed at small and zero applied magnetic fields where the ground state is disordered, there is a broad distribution of neutron response functions indicated by large values of $\xi$. As the polarized $|\hspace{-2pt}\uparrow \uparrow \uparrow \rangle$ phase is entered, the molecular field increases indicative of the formation of the polarized and spatially correlated $|\hspace{-2pt}\uparrow \uparrow \uparrow \rangle$ phase. It is interesting to note that full saturation of the molecular field is not onset immediately above the critical applied field of $\mu_{0}H_{c}\sim$ 1.6 T, and gradually forms with increased field, indicative of the second-order nature of the critical point.  Concomitantly, the parameter $\xi$ which is related to the distribution of neutron response functions $\kappa$ decreases indicative of a narrowing of the distribution of responses.

Our parametrization agrees with expectations based on the statics reported previously from diffraction and susceptibility.  While magnetic order is observed at zero field, it is only a fraction of the expected total based on observations in other cobalt oxide compounds.  This is indicative of a large fluctuating moment, corroborated by the low temperature $\mu_{0}H$=0 T response that we observed in Fig. \ref{figure_4} $(e)$.  This largely disordered ``liquid" like phase is parameterized by a broad range of localized response functions in our analysis.    

While the excitation spectrum at $\mu_{0}H$=0 T is broadened in both momentum and energy and lacks the strong dispersive nature of the excitations at high fields, it is less complex than intermediate fields in the $|\hspace{-2pt}\uparrow\uparrow\downarrow\rangle$ phase where a mixture of relaxational components near the magnetic zone center and dispersionless features were observed.  This spectrum is noted to host more efficient cooling than in the intermediate $|\hspace{-2pt}\uparrow\uparrow\downarrow\rangle$ phase.

NBCP displays efficient cooling through adiabatic demagnetization.  Based on our mapping and parameterization of the magnetic response discussed above, there are several reasons for this.  The primary reason is a ground state Kramer's doublet originating from the single-ion physics of Co$^{2+}$ which can only be split in the presence of a time reversal symmetry breaking magnetic field. Thus, Kramers doublets cannot be split in a time reversal symmetry perserving field such as a crystalline electric field.  Splitting for NBCP occurs for magnetic fields above $\mu_{0}H_{c}$=1.6 T.  Normally for field induced critical points defining a low-field magnetically ordered phase, the Kramers doublet would split again at lower fields below the critical magnetic field characterized by an opening of an excitation gap.~\cite{Magarino77:23,Ruegg08:100,Coldea10:327}

This is noted in the $|\hspace{-4pt}\uparrow\uparrow\downarrow\rangle$ phase, where an effective splitting seen in Figs. \ref{figure_3} $(e,f)$ causes a rise in temperature. We note that owing to disorder in this intermediate field phase, there are excitations which are dispersionless indicative of Co$^{2+}$ sites which lack any molecular field or coupling and increases the entropy in this field range that causes heating.  Meanwhile, at $\mu_{0}H$= 0 T the liquid like ground state preserves the Kramers ground state degeneracy and hence cools the sample.  The two essential components for efficient cooling in NBCP are therefore the single-ion physics that provides the basis for the Kramers doublets and frustrated spin interactions preventing full spatially long-range magnetic order at zero field.

In summary we have applied high resolution neutron spectroscopy to study the correlated Kramers doublet magnetism near the field induced critical point $\mu_{0}H_{c}\sim$1.6 T.  We have parameterzied the excitations in terms of a ground state $j_{eff}={1\over2}$ doublets coupled via the mean field random phase approximation.  We have included a broadening parameter to account for a distribution of responses owing to the inherently disordered ground state.  We have also mapped out the excitations in the low-field disordered and $|\hspace{-2pt}\uparrow\uparrow\downarrow\rangle$ phases.    


\begin{thebibliography}{45}%
	\makeatletter
	\providecommand \@ifxundefined [1]{%
		\@ifx{#1\undefined}
	}%
	\providecommand \@ifnum [1]{%
		\ifnum #1\expandafter \@firstoftwo
		\else \expandafter \@secondoftwo
		\fi
	}%
	\providecommand \@ifx [1]{%
		\ifx #1\expandafter \@firstoftwo
		\else \expandafter \@secondoftwo
		\fi
	}%
	\providecommand \natexlab [1]{#1}%
	\providecommand \enquote  [1]{``#1''}%
	\providecommand \bibnamefont  [1]{#1}%
	\providecommand \bibfnamefont [1]{#1}%
	\providecommand \citenamefont [1]{#1}%
	\providecommand \href@noop [0]{\@secondoftwo}%
	\providecommand \href [0]{\begingroup \@sanitize@url \@href}%
	\providecommand \@href[1]{\@@startlink{#1}\@@href}%
	\providecommand \@@href[1]{\endgroup#1\@@endlink}%
	\providecommand \@sanitize@url [0]{\catcode `\\12\catcode `\$12\catcode
		`\&12\catcode `\#12\catcode `\^12\catcode `\_12\catcode `\%12\relax}%
	\providecommand \@@startlink[1]{}%
	\providecommand \@@endlink[0]{}%
	\providecommand \url  [0]{\begingroup\@sanitize@url \@url }%
	\providecommand \@url [1]{\endgroup\@href {#1}{\urlprefix }}%
	\providecommand \urlprefix  [0]{URL }%
	\providecommand \Eprint [0]{\href }%
	\providecommand \doibase [0]{http://dx.doi.org/}%
	\providecommand \selectlanguage [0]{\@gobble}%
	\providecommand \bibinfo  [0]{\@secondoftwo}%
	\providecommand \bibfield  [0]{\@secondoftwo}%
	\providecommand \translation [1]{[#1]}%
	\providecommand \BibitemOpen [0]{}%
	\providecommand \bibitemStop [0]{}%
	\providecommand \bibitemNoStop [0]{.\EOS\space}%
	\providecommand \EOS [0]{\spacefactor3000\relax}%
	\providecommand \BibitemShut  [1]{\csname bibitem#1\endcsname}%
	\let\auto@bib@innerbib\@empty
	\bibitem [{\citenamefont {Xiang}\ \emph {et~al.}(2024)\citenamefont {Xiang},
		\citenamefont {Zhang}, \citenamefont {Gao}, \citenamefont {Schmidt},
		\citenamefont {Schmalzl}, \citenamefont {Wang}, \citenamefont {Li},
		\citenamefont {Xi}, \citenamefont {Liu}, \citenamefont {Jin} \emph
		{et~al.}}]{Xiang24:625}%
	\BibitemOpen
	\bibfield  {author} {\bibinfo {author} {\bibfnamefont {Junsen}\ \bibnamefont
			{Xiang}}, \bibinfo {author} {\bibfnamefont {Chuandi}\ \bibnamefont {Zhang}},
		\bibinfo {author} {\bibfnamefont {Yuan}\ \bibnamefont {Gao}}, \bibinfo
		{author} {\bibfnamefont {Wolfgang}\ \bibnamefont {Schmidt}}, \bibinfo
		{author} {\bibfnamefont {Karin}\ \bibnamefont {Schmalzl}}, \bibinfo {author}
		{\bibfnamefont {Chin-Wei}\ \bibnamefont {Wang}}, \bibinfo {author}
		{\bibfnamefont {Bo}~\bibnamefont {Li}}, \bibinfo {author} {\bibfnamefont
			{Ning}\ \bibnamefont {Xi}}, \bibinfo {author} {\bibfnamefont {Xin-Yang}\
			\bibnamefont {Liu}}, \bibinfo {author} {\bibfnamefont {Hai}\ \bibnamefont
			{Jin}},  \emph {et~al.},\ }\bibfield  {title} {\enquote {\bibinfo {title}
			{Giant magnetocaloric effect in spin supersolid candidate
				\textsc{N}a$_{2}$\textsc{B}a\textsc{C}o(\textsc{PO}$_{4}$)$_{2}$},}\ }\href
	{\doibase 10.1038/s41586-023-06885-w} {\bibfield  {journal} {\bibinfo
			{journal} {Nature}\ }\textbf {\bibinfo {volume} {625}},\ \bibinfo {pages}
		{270} (\bibinfo {year} {2024})}\BibitemShut {NoStop}%
	\bibitem [{\citenamefont {Collins}\ and\ \citenamefont
		{Petrenko}(1997)}]{Collins97:75}%
	\BibitemOpen
	\bibfield  {author} {\bibinfo {author} {\bibfnamefont {M.~F.}\ \bibnamefont
			{Collins}}\ and\ \bibinfo {author} {\bibfnamefont {O.~A.}\ \bibnamefont
			{Petrenko}},\ }\bibfield  {title} {\enquote {\bibinfo {title} {Triangular
				antiferromagnets},}\ }\href {\doibase 10.1139/cjp-75-9-605} {\bibfield
		{journal} {\bibinfo  {journal} {Can. J. Phys.}\ }\textbf {\bibinfo {volume}
			{75}},\ \bibinfo {pages} {605} (\bibinfo {year} {1997})}\BibitemShut
	{NoStop}%
	\bibitem [{\citenamefont {Tsunetsugu}\ and\ \citenamefont
		{Arikawa}(2006)}]{Tsunetsugu06:75}%
	\BibitemOpen
	\bibfield  {author} {\bibinfo {author} {\bibfnamefont {H.}~\bibnamefont
			{Tsunetsugu}}\ and\ \bibinfo {author} {\bibfnamefont {M.}~\bibnamefont
			{Arikawa}},\ }\bibfield  {title} {\enquote {\bibinfo {title} {Spin nematic
				phase in $\textsc{S}$=1 triangular antiferromagnets},}\ }\href {\doibase
		10.1143/JPSJ.75.083701} {\bibfield  {journal} {\bibinfo  {journal} {J. Phys.
				Soc. Jpn.}\ }\textbf {\bibinfo {volume} {75}},\ \bibinfo {pages} {083701}
		(\bibinfo {year} {2006})}\BibitemShut {NoStop}%
	\bibitem [{\citenamefont {Stoudenmire}\ \emph {et~al.}(2009)\citenamefont
		{Stoudenmire}, \citenamefont {Trebst},\ and\ \citenamefont
		{Balents}}]{Stoudenmire09:79}%
	\BibitemOpen
	\bibfield  {author} {\bibinfo {author} {\bibfnamefont {E.~M.}\ \bibnamefont
			{Stoudenmire}}, \bibinfo {author} {\bibfnamefont {Simon}\ \bibnamefont
			{Trebst}}, \ and\ \bibinfo {author} {\bibfnamefont {Leon}\ \bibnamefont
			{Balents}},\ }\bibfield  {title} {\enquote {\bibinfo {title} {Quadrupolar
				correlations and spin freezing in $\textsc{S}=1$ triangular lattice
				antiferromagnets},}\ }\href {\doibase 10.1103/PhysRevB.79.214436} {\bibfield
		{journal} {\bibinfo  {journal} {Phys. Rev. B}\ }\textbf {\bibinfo {volume}
			{79}},\ \bibinfo {pages} {214436} (\bibinfo {year} {2009})}\BibitemShut
	{NoStop}%
	\bibitem [{\citenamefont {Anderson}(1973)}]{Anderson73:8}%
	\BibitemOpen
	\bibfield  {author} {\bibinfo {author} {\bibfnamefont {P.~W.}\ \bibnamefont
			{Anderson}},\ }\bibfield  {title} {\enquote {\bibinfo {title} {Resonating
				valence bonds: A new kind of insulator},}\ }\href {\doibase
		10.1016/0025-5408(73)90167-0} {\bibfield  {journal} {\bibinfo  {journal}
			{Mater. Res. Bull.}\ }\textbf {\bibinfo {volume} {8}},\ \bibinfo {pages}
		{153} (\bibinfo {year} {1973})}\BibitemShut {NoStop}%
	\bibitem [{\citenamefont {Wessel}\ and\ \citenamefont
		{Troyer}(2005)}]{Wessel05:95}%
	\BibitemOpen
	\bibfield  {author} {\bibinfo {author} {\bibfnamefont {Stefan}\ \bibnamefont
			{Wessel}}\ and\ \bibinfo {author} {\bibfnamefont {Matthias}\ \bibnamefont
			{Troyer}},\ }\bibfield  {title} {\enquote {\bibinfo {title} {Supersolid
				hard-core bosons on the triangular lattice},}\ }\href {\doibase
		10.1103/PhysRevLett.95.127205} {\bibfield  {journal} {\bibinfo  {journal}
			{Phys. Rev. Lett.}\ }\textbf {\bibinfo {volume} {95}},\ \bibinfo {pages}
		{127205} (\bibinfo {year} {2005})}\BibitemShut {NoStop}%
	\bibitem [{\citenamefont {Melko}\ \emph {et~al.}(2005)\citenamefont {Melko},
		\citenamefont {Paramekanti}, \citenamefont {Burkov}, \citenamefont
		{Vishwanath}, \citenamefont {Sheng},\ and\ \citenamefont
		{Balents}}]{Melko05:95}%
	\BibitemOpen
	\bibfield  {author} {\bibinfo {author} {\bibfnamefont {R.~G.}\ \bibnamefont
			{Melko}}, \bibinfo {author} {\bibfnamefont {A.}~\bibnamefont {Paramekanti}},
		\bibinfo {author} {\bibfnamefont {A.~A.}\ \bibnamefont {Burkov}}, \bibinfo
		{author} {\bibfnamefont {A.}~\bibnamefont {Vishwanath}}, \bibinfo {author}
		{\bibfnamefont {D.~N.}\ \bibnamefont {Sheng}}, \ and\ \bibinfo {author}
		{\bibfnamefont {L.}~\bibnamefont {Balents}},\ }\bibfield  {title} {\enquote
		{\bibinfo {title} {Supersolid order from disorder: Hard-core bosons on the
				triangular lattice},}\ }\href {\doibase 10.1103/PhysRevLett.95.127207}
	{\bibfield  {journal} {\bibinfo  {journal} {Phys. Rev. Lett.}\ }\textbf
		{\bibinfo {volume} {95}},\ \bibinfo {pages} {127207} (\bibinfo {year}
		{2005})}\BibitemShut {NoStop}%
	\bibitem [{\citenamefont {Heidarian}\ and\ \citenamefont
		{Damle}(2005)}]{Heidarian05:95}%
	\BibitemOpen
	\bibfield  {author} {\bibinfo {author} {\bibfnamefont {Dariush}\ \bibnamefont
			{Heidarian}}\ and\ \bibinfo {author} {\bibfnamefont {Kedar}\ \bibnamefont
			{Damle}},\ }\bibfield  {title} {\enquote {\bibinfo {title} {Persistent
				supersolid phase of hard-core bosons on the triangular lattice},}\ }\href
	{\doibase 10.1103/PhysRevLett.95.127206} {\bibfield  {journal} {\bibinfo
			{journal} {Phys. Rev. Lett.}\ }\textbf {\bibinfo {volume} {95}},\ \bibinfo
		{pages} {127206} (\bibinfo {year} {2005})}\BibitemShut {NoStop}%
	\bibitem [{\citenamefont {Boninsegni}\ and\ \citenamefont
		{Prokof'ev}(2005)}]{Boninsegni05:95}%
	\BibitemOpen
	\bibfield  {author} {\bibinfo {author} {\bibfnamefont {Massimo}\ \bibnamefont
			{Boninsegni}}\ and\ \bibinfo {author} {\bibfnamefont {Nikolay}\ \bibnamefont
			{Prokof'ev}},\ }\bibfield  {title} {\enquote {\bibinfo {title} {Supersolid
				phase of hard-core bosons on a triangular lattice},}\ }\href {\doibase
		10.1103/PhysRevLett.95.237204} {\bibfield  {journal} {\bibinfo  {journal}
			{Phys. Rev. Lett.}\ }\textbf {\bibinfo {volume} {95}},\ \bibinfo {pages}
		{237204} (\bibinfo {year} {2005})}\BibitemShut {NoStop}%
	\bibitem [{\citenamefont {Heidarian}\ and\ \citenamefont
		{Paramekanti}(2010)}]{Heidarian10:104}%
	\BibitemOpen
	\bibfield  {author} {\bibinfo {author} {\bibfnamefont {Dariush}\ \bibnamefont
			{Heidarian}}\ and\ \bibinfo {author} {\bibfnamefont {Arun}\ \bibnamefont
			{Paramekanti}},\ }\bibfield  {title} {\enquote {\bibinfo {title}
			{Supersolidity in the triangular lattice spin-$1/2$ $\textsc{XXZ}$ model: A
				variational perspective},}\ }\href {\doibase 10.1103/PhysRevLett.104.015301}
	{\bibfield  {journal} {\bibinfo  {journal} {Phys. Rev. Lett.}\ }\textbf
		{\bibinfo {volume} {104}},\ \bibinfo {pages} {015301} (\bibinfo {year}
		{2010})}\BibitemShut {NoStop}%
	\bibitem [{\citenamefont {Wang}\ \emph {et~al.}(2009)\citenamefont {Wang},
		\citenamefont {Pollmann},\ and\ \citenamefont {Vishwanath}}]{Wang09:102}%
	\BibitemOpen
	\bibfield  {author} {\bibinfo {author} {\bibfnamefont {Fa}~\bibnamefont
			{Wang}}, \bibinfo {author} {\bibfnamefont {Frank}\ \bibnamefont {Pollmann}},
		\ and\ \bibinfo {author} {\bibfnamefont {Ashvin}\ \bibnamefont
			{Vishwanath}},\ }\bibfield  {title} {\enquote {\bibinfo {title} {Extended
				supersolid phase of frustrated hard-core bosons on a triangular lattice},}\
	}\href {\doibase 10.1103/PhysRevLett.102.017203} {\bibfield  {journal}
		{\bibinfo  {journal} {Phys. Rev. Lett.}\ }\textbf {\bibinfo {volume} {102}},\
		\bibinfo {pages} {017203} (\bibinfo {year} {2009})}\BibitemShut {NoStop}%
	\bibitem [{\citenamefont {Jiang}\ \emph {et~al.}(2009)\citenamefont {Jiang},
		\citenamefont {Weng}, \citenamefont {Weng}, \citenamefont {Sheng},\ and\
		\citenamefont {Balents}}]{Jiang09:79}%
	\BibitemOpen
	\bibfield  {author} {\bibinfo {author} {\bibfnamefont {H.~C.}\ \bibnamefont
			{Jiang}}, \bibinfo {author} {\bibfnamefont {M.~Q.}\ \bibnamefont {Weng}},
		\bibinfo {author} {\bibfnamefont {Z.~Y.}\ \bibnamefont {Weng}}, \bibinfo
		{author} {\bibfnamefont {D.~N.}\ \bibnamefont {Sheng}}, \ and\ \bibinfo
		{author} {\bibfnamefont {L.}~\bibnamefont {Balents}},\ }\bibfield  {title}
	{\enquote {\bibinfo {title} {Supersolid order of frustrated hard-core bosons
				in a triangular lattice system},}\ }\href {\doibase
		10.1103/PhysRevB.79.020409} {\bibfield  {journal} {\bibinfo  {journal} {Phys.
				Rev. B}\ }\textbf {\bibinfo {volume} {79}},\ \bibinfo {pages} {020409(R)}
		(\bibinfo {year} {2009})}\BibitemShut {NoStop}%
	\bibitem [{\citenamefont {Kim}\ and\ \citenamefont {Chan}(2004)}]{Kim04:427}%
	\BibitemOpen
	\bibfield  {author} {\bibinfo {author} {\bibfnamefont {E.}~\bibnamefont
			{Kim}}\ and\ \bibinfo {author} {\bibfnamefont {M.~H.~W.}\ \bibnamefont
			{Chan}},\ }\bibfield  {title} {\enquote {\bibinfo {title} {Probable
				observation of a supersolid helium phase},}\ }\href {\doibase
		10.1038/nature02220} {\bibfield  {journal} {\bibinfo  {journal} {Nature}\
		}\textbf {\bibinfo {volume} {427}},\ \bibinfo {pages} {225} (\bibinfo {year}
		{2004})}\BibitemShut {NoStop}%
	\bibitem [{\citenamefont {Gao}\ \emph {et~al.}(2022)\citenamefont {Gao},
		\citenamefont {Fan}, \citenamefont {Li}, \citenamefont {Yang}, \citenamefont
		{Zeng}, \citenamefont {Sheng}, \citenamefont {Zhong}, \citenamefont {Qi},
		\citenamefont {Wan},\ and\ \citenamefont {Li}}]{Gao22:89}%
	\BibitemOpen
	\bibfield  {author} {\bibinfo {author} {\bibfnamefont {Y.}~\bibnamefont
			{Gao}}, \bibinfo {author} {\bibfnamefont {Y-C}\ \bibnamefont {Fan}}, \bibinfo
		{author} {\bibfnamefont {H.}~\bibnamefont {Li}}, \bibinfo {author}
		{\bibfnamefont {F.}~\bibnamefont {Yang}}, \bibinfo {author} {\bibfnamefont
			{X.-T}\ \bibnamefont {Zeng}}, \bibinfo {author} {\bibfnamefont {X.-L.}\
			\bibnamefont {Sheng}}, \bibinfo {author} {\bibfnamefont {R.}~\bibnamefont
			{Zhong}}, \bibinfo {author} {\bibfnamefont {Y.}~\bibnamefont {Qi}}, \bibinfo
		{author} {\bibfnamefont {Y.}~\bibnamefont {Wan}}, \ and\ \bibinfo {author}
		{\bibfnamefont {W.}~\bibnamefont {Li}},\ }\bibfield  {title} {\enquote
		{\bibinfo {title} {Spin supersolidity in nearly ideal easy-axis triangular
				quantum antiferromagnet
				\textsc{N}a$_2$\textsc{B}a\textsc{C}o(\textsc{PO}$_4$)$_2$},}\ }\href
	{\doibase 10.1038/s41535-022-00500-3} {\bibfield  {journal} {\bibinfo
			{journal} {npj Quantum Mater.}\ }\textbf {\bibinfo {volume} {7}},\ \bibinfo
		{pages} {89} (\bibinfo {year} {2022})}\BibitemShut {NoStop}%
	\bibitem [{\citenamefont {Stryjewski}\ and\ \citenamefont
		{Giordano}(1977)}]{Stryjewski77:26}%
	\BibitemOpen
	\bibfield  {author} {\bibinfo {author} {\bibfnamefont {E.}~\bibnamefont
			{Stryjewski}}\ and\ \bibinfo {author} {\bibfnamefont {N.}~\bibnamefont
			{Giordano}},\ }\bibfield  {title} {\enquote {\bibinfo {title}
			{Metamagnetism},}\ }\href {\doibase 10.1080/00018737700101433} {\bibfield
		{journal} {\bibinfo  {journal} {Adv. Phys.}\ }\textbf {\bibinfo {volume}
			{26}},\ \bibinfo {pages} {487} (\bibinfo {year} {1977})}\BibitemShut
	{NoStop}%
	\bibitem [{\citenamefont {Zhang}\ \emph {et~al.}(2024)\citenamefont {Zhang},
		\citenamefont {Xiang}, \citenamefont {Zhu}, \citenamefont {Wu}, \citenamefont
		{Zhang}, \citenamefont {Xu}, \citenamefont {Yin}, \citenamefont {Sun},
		\citenamefont {Li}, \citenamefont {Su},\ and\ \citenamefont
		{Jin}}]{Zhang24:8}%
	\BibitemOpen
	\bibfield  {author} {\bibinfo {author} {\bibfnamefont {Chuandi}\ \bibnamefont
			{Zhang}}, \bibinfo {author} {\bibfnamefont {Junsen}\ \bibnamefont {Xiang}},
		\bibinfo {author} {\bibfnamefont {Quanliang}\ \bibnamefont {Zhu}}, \bibinfo
		{author} {\bibfnamefont {Longfei}\ \bibnamefont {Wu}}, \bibinfo {author}
		{\bibfnamefont {Shanfeng}\ \bibnamefont {Zhang}}, \bibinfo {author}
		{\bibfnamefont {Juping}\ \bibnamefont {Xu}}, \bibinfo {author} {\bibfnamefont
			{Wen}\ \bibnamefont {Yin}}, \bibinfo {author} {\bibfnamefont {Peijie}\
			\bibnamefont {Sun}}, \bibinfo {author} {\bibfnamefont {Wei}\ \bibnamefont
			{Li}}, \bibinfo {author} {\bibfnamefont {Gang}\ \bibnamefont {Su}}, \ and\
		\bibinfo {author} {\bibfnamefont {Wentao}\ \bibnamefont {Jin}},\ }\bibfield
	{title} {\enquote {\bibinfo {title} {Structural, magnetic, and magnetocaloric
				properties of triangular-lattice transition-metal phosphates},}\ }\href
	{\doibase 10.1103/PhysRevMaterials.8.044409} {\bibfield  {journal} {\bibinfo
			{journal} {Phys. Rev. Mater.}\ }\textbf {\bibinfo {volume} {8}},\ \bibinfo
		{pages} {044409} (\bibinfo {year} {2024})}\BibitemShut {NoStop}%
	\bibitem [{\citenamefont {Sheng}\ \emph {et~al.}(2022)\citenamefont {Sheng},
		\citenamefont {Wang}, \citenamefont {Candini}, \citenamefont {Jiang},
		\citenamefont {Huang}, \citenamefont {Xi}, \citenamefont {Zhao},
		\citenamefont {Ge}, \citenamefont {Zhao}, \citenamefont {Fu} \emph
		{et~al.}}]{Sheng22:119}%
	\BibitemOpen
	\bibfield  {author} {\bibinfo {author} {\bibfnamefont {Jieming}\ \bibnamefont
			{Sheng}}, \bibinfo {author} {\bibfnamefont {Le}~\bibnamefont {Wang}},
		\bibinfo {author} {\bibfnamefont {Andrea}\ \bibnamefont {Candini}}, \bibinfo
		{author} {\bibfnamefont {Wenrui}\ \bibnamefont {Jiang}}, \bibinfo {author}
		{\bibfnamefont {Lianglong}\ \bibnamefont {Huang}}, \bibinfo {author}
		{\bibfnamefont {Bin}\ \bibnamefont {Xi}}, \bibinfo {author} {\bibfnamefont
			{Jize}\ \bibnamefont {Zhao}}, \bibinfo {author} {\bibfnamefont {Han}\
			\bibnamefont {Ge}}, \bibinfo {author} {\bibfnamefont {Nan}\ \bibnamefont
			{Zhao}}, \bibinfo {author} {\bibfnamefont {Ying}\ \bibnamefont {Fu}},  \emph
		{et~al.},\ }\bibfield  {title} {\enquote {\bibinfo {title} {Two-dimensional
				quantum universality in the spin-1/2 triangular-lattice quantum
				antiferromagnet
				\textsc{N}a$_{2}$\textsc{B}a\textsc{C}o(\textsc{PO}$_{4}$)$_{2}$},}\ }\href
	{\doibase 10.1073/pnas.2211193119} {\bibfield  {journal} {\bibinfo  {journal}
			{Proc. Natl. Acad. Sci. USA}\ }\textbf {\bibinfo {volume} {119}},\ \bibinfo
		{pages} {e2211193119} (\bibinfo {year} {2022})}\BibitemShut {NoStop}%
	\bibitem [{\citenamefont {Lee}\ \emph {et~al.}(2021)\citenamefont {Lee},
		\citenamefont {Lee}, \citenamefont {Berlie}, \citenamefont {Hillier},
		\citenamefont {Adroja}, \citenamefont {Zhong}, \citenamefont {Cava},
		\citenamefont {Jang},\ and\ \citenamefont {Choi}}]{Lee21:103}%
	\BibitemOpen
	\bibfield  {author} {\bibinfo {author} {\bibfnamefont {S.}~\bibnamefont
			{Lee}}, \bibinfo {author} {\bibfnamefont {C.~H.}\ \bibnamefont {Lee}},
		\bibinfo {author} {\bibfnamefont {A.}~\bibnamefont {Berlie}}, \bibinfo
		{author} {\bibfnamefont {A.~D.}\ \bibnamefont {Hillier}}, \bibinfo {author}
		{\bibfnamefont {Devashibhai~T.}\ \bibnamefont {Adroja}}, \bibinfo {author}
		{\bibfnamefont {Ruidan}\ \bibnamefont {Zhong}}, \bibinfo {author}
		{\bibfnamefont {R.~J.}\ \bibnamefont {Cava}}, \bibinfo {author}
		{\bibfnamefont {Z.~H.}\ \bibnamefont {Jang}}, \ and\ \bibinfo {author}
		{\bibfnamefont {K.-Y.}\ \bibnamefont {Choi}},\ }\bibfield  {title} {\enquote
		{\bibinfo {title} {Temporal and field evolution of spin excitations in the
				disorder-free triangular antiferromagnet
				\textsc{N}a$_{2}$\textsc{B}a\textsc{C}o(\textsc{P}\textsc{O}$_{4}$)$_{2}$},}\
	}\href {\doibase 10.1103/PhysRevB.103.024413} {\bibfield  {journal} {\bibinfo
			{journal} {Phys. Rev. B}\ }\textbf {\bibinfo {volume} {103}},\ \bibinfo
		{pages} {024413} (\bibinfo {year} {2021})}\BibitemShut {NoStop}%
	\bibitem [{\citenamefont {Zhong}\ \emph {et~al.}(2019)\citenamefont {Zhong},
		\citenamefont {Guo}, \citenamefont {Xu}, \citenamefont {Xu},\ and\
		\citenamefont {Cava}}]{Zhong19:116}%
	\BibitemOpen
	\bibfield  {author} {\bibinfo {author} {\bibfnamefont {R.}~\bibnamefont
			{Zhong}}, \bibinfo {author} {\bibfnamefont {S.}~\bibnamefont {Guo}}, \bibinfo
		{author} {\bibfnamefont {G.}~\bibnamefont {Xu}}, \bibinfo {author}
		{\bibfnamefont {Z.}~\bibnamefont {Xu}}, \ and\ \bibinfo {author}
		{\bibfnamefont {R.~J.}\ \bibnamefont {Cava}},\ }\bibfield  {title} {\enquote
		{\bibinfo {title} {Strong quantum fluctuations in a quantum spin liquid
				candidate with a \textsc{C}o-based triangular lattice},}\ }\href {\doibase
		10.1073/pnas.1906483116} {\bibfield  {journal} {\bibinfo  {journal} {Proc.
				Natl. Acad. Sci. USA}\ }\textbf {\bibinfo {volume} {116}},\ \bibinfo {pages}
		{14505} (\bibinfo {year} {2019})}\BibitemShut {NoStop}%
	\bibitem [{\citenamefont {Li}\ \emph {et~al.}(2020)\citenamefont {Li},
		\citenamefont {Huang}, \citenamefont {Yue}, \citenamefont {Chu},
		\citenamefont {Chen}, \citenamefont {Choi}, \citenamefont {Zhao},
		\citenamefont {Zhou},\ and\ \citenamefont {Sun}}]{Li20:11}%
	\BibitemOpen
	\bibfield  {author} {\bibinfo {author} {\bibfnamefont {N.}~\bibnamefont
			{Li}}, \bibinfo {author} {\bibfnamefont {Q.}~\bibnamefont {Huang}}, \bibinfo
		{author} {\bibfnamefont {X.~Y.}\ \bibnamefont {Yue}}, \bibinfo {author}
		{\bibfnamefont {W.~J.}\ \bibnamefont {Chu}}, \bibinfo {author} {\bibfnamefont
			{Q.}~\bibnamefont {Chen}}, \bibinfo {author} {\bibfnamefont {E.~S.}\
			\bibnamefont {Choi}}, \bibinfo {author} {\bibfnamefont {X.}~\bibnamefont
			{Zhao}}, \bibinfo {author} {\bibfnamefont {H.~D.}\ \bibnamefont {Zhou}}, \
		and\ \bibinfo {author} {\bibfnamefont {X.~F.}\ \bibnamefont {Sun}},\
	}\bibfield  {title} {\enquote {\bibinfo {title} {Possible itinerant
				excitations and quantum spin state transitions in the effective spin-${1\over
					2}$ triangular-lattice antiferromagnet
				\textsc{N}a$_{2}$\textsc{B}a\textsc{C}o(\textsc{P}\textsc{O}$_{4}$)$_{2}$},}\
	}\href {\doibase 10.1038/s41467-020-18041-3} {\bibfield  {journal} {\bibinfo
			{journal} {Nat. Commun}\ }\textbf {\bibinfo {volume} {11}},\ \bibinfo {pages}
		{4216} (\bibinfo {year} {2020})}\BibitemShut {NoStop}%
	\bibitem [{\citenamefont {Gao}\ \emph {et~al.}(2024)\citenamefont {Gao},
		\citenamefont {Zhang}, \citenamefont {Xiang}, \citenamefont {Yu},
		\citenamefont {Lu}, \citenamefont {Sun}, \citenamefont {Jin}, \citenamefont
		{Su},\ and\ \citenamefont {Li}}]{Gao24:110}%
	\BibitemOpen
	\bibfield  {author} {\bibinfo {author} {\bibfnamefont {Yuan}\ \bibnamefont
			{Gao}}, \bibinfo {author} {\bibfnamefont {Chuandi}\ \bibnamefont {Zhang}},
		\bibinfo {author} {\bibfnamefont {Junsen}\ \bibnamefont {Xiang}}, \bibinfo
		{author} {\bibfnamefont {Dehong}\ \bibnamefont {Yu}}, \bibinfo {author}
		{\bibfnamefont {Xingye}\ \bibnamefont {Lu}}, \bibinfo {author} {\bibfnamefont
			{Peijie}\ \bibnamefont {Sun}}, \bibinfo {author} {\bibfnamefont {Wentao}\
			\bibnamefont {Jin}}, \bibinfo {author} {\bibfnamefont {Gang}\ \bibnamefont
			{Su}}, \ and\ \bibinfo {author} {\bibfnamefont {Wei}\ \bibnamefont {Li}},\
	}\bibfield  {title} {\enquote {\bibinfo {title} {Double magnon-roton
				excitations in the triangular-lattice spin supersolid},}\ }\href {\doibase
		10.1103/PhysRevB.110.214408} {\bibfield  {journal} {\bibinfo  {journal}
			{Phys. Rev. B}\ }\textbf {\bibinfo {volume} {110}},\ \bibinfo {pages}
		{214408} (\bibinfo {year} {2024})}\BibitemShut {NoStop}%
	\bibitem [{\citenamefont {Sakurai}\ \emph {et~al.}(1968)\citenamefont
		{Sakurai}, \citenamefont {Buyers}, \citenamefont {Cowley},\ and\
		\citenamefont {Dolling}}]{Sakurai68:167}%
	\BibitemOpen
	\bibfield  {author} {\bibinfo {author} {\bibfnamefont {J.}~\bibnamefont
			{Sakurai}}, \bibinfo {author} {\bibfnamefont {W.~J.~L.}\ \bibnamefont
			{Buyers}}, \bibinfo {author} {\bibfnamefont {R.~A.}\ \bibnamefont {Cowley}},
		\ and\ \bibinfo {author} {\bibfnamefont {G.}~\bibnamefont {Dolling}},\
	}\bibfield  {title} {\enquote {\bibinfo {title} {Crystal dynamics and
				magnetic excitations in cobaltous oxide},}\ }\href {\doibase
		10.1103/PhysRev.167.510} {\bibfield  {journal} {\bibinfo  {journal} {Phys.
				Rev.}\ }\textbf {\bibinfo {volume} {167}},\ \bibinfo {pages} {510} (\bibinfo
		{year} {1968})}\BibitemShut {NoStop}%
	\bibitem [{\citenamefont {Cowley}\ \emph {et~al.}(2013)\citenamefont {Cowley},
		\citenamefont {Buyers}, \citenamefont {Stock}, \citenamefont {Yamani},
		\citenamefont {Frost}, \citenamefont {Taylor},\ and\ \citenamefont
		{Prabhakaran}}]{Cowley13:88}%
	\BibitemOpen
	\bibfield  {author} {\bibinfo {author} {\bibfnamefont {R.~A.}\ \bibnamefont
			{Cowley}}, \bibinfo {author} {\bibfnamefont {W.~J.~L.}\ \bibnamefont
			{Buyers}}, \bibinfo {author} {\bibfnamefont {C.}~\bibnamefont {Stock}},
		\bibinfo {author} {\bibfnamefont {Z.}~\bibnamefont {Yamani}}, \bibinfo
		{author} {\bibfnamefont {C.}~\bibnamefont {Frost}}, \bibinfo {author}
		{\bibfnamefont {J.~W.}\ \bibnamefont {Taylor}}, \ and\ \bibinfo {author}
		{\bibfnamefont {D.}~\bibnamefont {Prabhakaran}},\ }\bibfield  {title}
	{\enquote {\bibinfo {title} {Neutron scattering investigation of the
				$d\ensuremath{-}d$ excitations below the mott gap of
				\textsc{C}o\textsc{O}},}\ }\href {\doibase 10.1103/PhysRevB.88.205117}
	{\bibfield  {journal} {\bibinfo  {journal} {Phys. Rev. B}\ }\textbf {\bibinfo
			{volume} {88}},\ \bibinfo {pages} {205117} (\bibinfo {year}
		{2013})}\BibitemShut {NoStop}%
	\bibitem [{\citenamefont {Sarte}\ \emph {et~al.}(2018)\citenamefont {Sarte},
		\citenamefont {Cowley}, \citenamefont {Rodriguez}, \citenamefont {Pachoud},
		\citenamefont {Le}, \citenamefont {Garc\'{\i}a-Sakai}, \citenamefont
		{Taylor}, \citenamefont {Frost}, \citenamefont {Prabhakaran}, \citenamefont
		{MacEwen}, \citenamefont {Kitada}, \citenamefont {Browne}, \citenamefont
		{Songvilay}, \citenamefont {Yamani}, \citenamefont {Buyers}, \citenamefont
		{Attfield},\ and\ \citenamefont {Stock}}]{Sarte18:98}%
	\BibitemOpen
	\bibfield  {author} {\bibinfo {author} {\bibfnamefont {P.~M.}\ \bibnamefont
			{Sarte}}, \bibinfo {author} {\bibfnamefont {R.~A.}\ \bibnamefont {Cowley}},
		\bibinfo {author} {\bibfnamefont {E.~E.}\ \bibnamefont {Rodriguez}}, \bibinfo
		{author} {\bibfnamefont {E.}~\bibnamefont {Pachoud}}, \bibinfo {author}
		{\bibfnamefont {D.}~\bibnamefont {Le}}, \bibinfo {author} {\bibfnamefont
			{V.}~\bibnamefont {Garc\'{\i}a-Sakai}}, \bibinfo {author} {\bibfnamefont
			{J.~W.}\ \bibnamefont {Taylor}}, \bibinfo {author} {\bibfnamefont {C.~D.}\
			\bibnamefont {Frost}}, \bibinfo {author} {\bibfnamefont {D.}~\bibnamefont
			{Prabhakaran}}, \bibinfo {author} {\bibfnamefont {C.}~\bibnamefont
			{MacEwen}}, \bibinfo {author} {\bibfnamefont {A.}~\bibnamefont {Kitada}},
		\bibinfo {author} {\bibfnamefont {A.~J.}\ \bibnamefont {Browne}}, \bibinfo
		{author} {\bibfnamefont {M.}~\bibnamefont {Songvilay}}, \bibinfo {author}
		{\bibfnamefont {Z.}~\bibnamefont {Yamani}}, \bibinfo {author} {\bibfnamefont
			{W.~J.~L.}\ \bibnamefont {Buyers}}, \bibinfo {author} {\bibfnamefont {J.~P.}\
			\bibnamefont {Attfield}}, \ and\ \bibinfo {author} {\bibfnamefont
			{C.}~\bibnamefont {Stock}},\ }\bibfield  {title} {\enquote {\bibinfo {title}
			{Disentangling orbital and spin exchange interactions for \textsc{C}o$^{2+}$
				on a rocksalt lattice},}\ }\href {\doibase 10.1103/PhysRevB.98.024415}
	{\bibfield  {journal} {\bibinfo  {journal} {Phys. Rev. B}\ }\textbf {\bibinfo
			{volume} {98}},\ \bibinfo {pages} {024415} (\bibinfo {year}
		{2018})}\BibitemShut {NoStop}%
	\bibitem [{\citenamefont {Sarte}\ \emph {et~al.}(2019)\citenamefont {Sarte},
		\citenamefont {Songvilay}, \citenamefont {Pachoud}, \citenamefont {Ewings},
		\citenamefont {Frost}, \citenamefont {Prabhakaran}, \citenamefont {Hong},
		\citenamefont {Browne}, \citenamefont {Yamani}, \citenamefont {Attfield},
		\citenamefont {Rodriguez}, \citenamefont {Wilson},\ and\ \citenamefont
		{Stock}}]{Sarte19:100}%
	\BibitemOpen
	\bibfield  {author} {\bibinfo {author} {\bibfnamefont {P.~M.}\ \bibnamefont
			{Sarte}}, \bibinfo {author} {\bibfnamefont {M.}~\bibnamefont {Songvilay}},
		\bibinfo {author} {\bibfnamefont {E.}~\bibnamefont {Pachoud}}, \bibinfo
		{author} {\bibfnamefont {R.~A.}\ \bibnamefont {Ewings}}, \bibinfo {author}
		{\bibfnamefont {C.~D.}\ \bibnamefont {Frost}}, \bibinfo {author}
		{\bibfnamefont {D.}~\bibnamefont {Prabhakaran}}, \bibinfo {author}
		{\bibfnamefont {K.~H.}\ \bibnamefont {Hong}}, \bibinfo {author}
		{\bibfnamefont {A.~J.}\ \bibnamefont {Browne}}, \bibinfo {author}
		{\bibfnamefont {Z.}~\bibnamefont {Yamani}}, \bibinfo {author} {\bibfnamefont
			{J.~P.}\ \bibnamefont {Attfield}}, \bibinfo {author} {\bibfnamefont {E.~E.}\
			\bibnamefont {Rodriguez}}, \bibinfo {author} {\bibfnamefont {S.~D.}\
			\bibnamefont {Wilson}}, \ and\ \bibinfo {author} {\bibfnamefont
			{C.}~\bibnamefont {Stock}},\ }\bibfield  {title} {\enquote {\bibinfo {title}
			{Spin-orbit excitons in \textsc{C}o\textsc{O}},}\ }\href {\doibase
		10.1103/PhysRevB.100.075143} {\bibfield  {journal} {\bibinfo  {journal}
			{Phys. Rev. B}\ }\textbf {\bibinfo {volume} {100}},\ \bibinfo {pages}
		{075143} (\bibinfo {year} {2019})}\BibitemShut {NoStop}%
	\bibitem [{\citenamefont {Sarte}\ \emph {et~al.}(2020)\citenamefont {Sarte},
		\citenamefont {Stock}, \citenamefont {Ortiz}, \citenamefont {Hong},\ and\
		\citenamefont {Wilson}}]{Sarte20:102}%
	\BibitemOpen
	\bibfield  {author} {\bibinfo {author} {\bibfnamefont {P.~M.}\ \bibnamefont
			{Sarte}}, \bibinfo {author} {\bibfnamefont {C.}~\bibnamefont {Stock}},
		\bibinfo {author} {\bibfnamefont {B.~R.}\ \bibnamefont {Ortiz}}, \bibinfo
		{author} {\bibfnamefont {K.~H.}\ \bibnamefont {Hong}}, \ and\ \bibinfo
		{author} {\bibfnamefont {S.~D.}\ \bibnamefont {Wilson}},\ }\bibfield  {title}
	{\enquote {\bibinfo {title} {Van vleck excitons in
				\textsc{C}a$_{2}$\textsc{R}u\textsc{O}$_{4}$},}\ }\href {\doibase
		10.1103/PhysRevB.102.245119} {\bibfield  {journal} {\bibinfo  {journal}
			{Phys. Rev. B}\ }\textbf {\bibinfo {volume} {102}},\ \bibinfo {pages}
		{245119} (\bibinfo {year} {2020})}\BibitemShut {NoStop}%
	\bibitem [{\citenamefont {Lane}\ \emph {et~al.}(2022)\citenamefont {Lane},
		\citenamefont {Songvilay}, \citenamefont {Ewings},\ and\ \citenamefont
		{Stock}}]{Lane22:106}%
	\BibitemOpen
	\bibfield  {author} {\bibinfo {author} {\bibfnamefont {H.}~\bibnamefont
			{Lane}}, \bibinfo {author} {\bibfnamefont {M.}~\bibnamefont {Songvilay}},
		\bibinfo {author} {\bibfnamefont {R.~A.}\ \bibnamefont {Ewings}}, \ and\
		\bibinfo {author} {\bibfnamefont {C.}~\bibnamefont {Stock}},\ }\bibfield
	{title} {\enquote {\bibinfo {title} {Excitonic transverse and amplitude
				fluctuations in noncollinear and charge-ordered
				\textsc{R}b\textsc{F}e$^{2+}$\textsc{F}e$^{3+}$\textsc{F}$_{6}$},}\ }\href
	{\doibase 10.1103/PhysRevB.106.054431} {\bibfield  {journal} {\bibinfo
			{journal} {Phys. Rev. B}\ }\textbf {\bibinfo {volume} {106}},\ \bibinfo
		{pages} {054431} (\bibinfo {year} {2022})}\BibitemShut {NoStop}%
	\bibitem [{\citenamefont {Lane}\ \emph {et~al.}(2023)\citenamefont {Lane},
		\citenamefont {Sarte}, \citenamefont {Guratinder}, \citenamefont
		{Arevalo-Lopez}, \citenamefont {Perry}, \citenamefont {Hunter}, \citenamefont
		{Weber}, \citenamefont {Roessli}, \citenamefont {Stunault}, \citenamefont
		{Su}, \citenamefont {Ewings}, \citenamefont {Wilson}, \citenamefont {B\"oni},
		\citenamefont {Attfield},\ and\ \citenamefont {Stock}}]{Lane23:5}%
	\BibitemOpen
	\bibfield  {author} {\bibinfo {author} {\bibfnamefont {H.}~\bibnamefont
			{Lane}}, \bibinfo {author} {\bibfnamefont {P.~M.}\ \bibnamefont {Sarte}},
		\bibinfo {author} {\bibfnamefont {K.}~\bibnamefont {Guratinder}}, \bibinfo
		{author} {\bibfnamefont {A.~M.}\ \bibnamefont {Arevalo-Lopez}}, \bibinfo
		{author} {\bibfnamefont {R.~S.}\ \bibnamefont {Perry}}, \bibinfo {author}
		{\bibfnamefont {E.~C.}\ \bibnamefont {Hunter}}, \bibinfo {author}
		{\bibfnamefont {T.}~\bibnamefont {Weber}}, \bibinfo {author} {\bibfnamefont
			{B.}~\bibnamefont {Roessli}}, \bibinfo {author} {\bibfnamefont
			{A.}~\bibnamefont {Stunault}}, \bibinfo {author} {\bibfnamefont
			{Y.}~\bibnamefont {Su}}, \bibinfo {author} {\bibfnamefont {R.~A.}\
			\bibnamefont {Ewings}}, \bibinfo {author} {\bibfnamefont {S.~D.}\
			\bibnamefont {Wilson}}, \bibinfo {author} {\bibfnamefont {P.}~\bibnamefont
			{B\"oni}}, \bibinfo {author} {\bibfnamefont {J.~P.}\ \bibnamefont
			{Attfield}}, \ and\ \bibinfo {author} {\bibfnamefont {C.}~\bibnamefont
			{Stock}},\ }\bibfield  {title} {\enquote {\bibinfo {title} {Spin-orbital
				correlations from complex orbital order in
				\textsc{M}g\textsc{V}$_{2}$\textsc{O}$_{4}$},}\ }\href {\doibase
		10.1103/PhysRevResearch.5.043146} {\bibfield  {journal} {\bibinfo  {journal}
			{Phys. Rev. Res.}\ }\textbf {\bibinfo {volume} {5}},\ \bibinfo {pages}
		{043146} (\bibinfo {year} {2023})}\BibitemShut {NoStop}%
	\bibitem [{\citenamefont {Lane}\ \emph {et~al.}(2021)\citenamefont {Lane},
		\citenamefont {Rodriguez}, \citenamefont {Walker}, \citenamefont
		{Niedermayer}, \citenamefont {Stuhr}, \citenamefont {Bewley}, \citenamefont
		{Voneshen}, \citenamefont {Green}, \citenamefont {Rodriguez-Rivera},
		\citenamefont {Fouquet}, \citenamefont {Cheong}, \citenamefont {Attfield},
		\citenamefont {Ewings},\ and\ \citenamefont {Stock}}]{Lane21:104}%
	\BibitemOpen
	\bibfield  {author} {\bibinfo {author} {\bibfnamefont {H.}~\bibnamefont
			{Lane}}, \bibinfo {author} {\bibfnamefont {E.~E.}\ \bibnamefont {Rodriguez}},
		\bibinfo {author} {\bibfnamefont {H.~C.}\ \bibnamefont {Walker}}, \bibinfo
		{author} {\bibfnamefont {Ch.}\ \bibnamefont {Niedermayer}}, \bibinfo {author}
		{\bibfnamefont {U.}~\bibnamefont {Stuhr}}, \bibinfo {author} {\bibfnamefont
			{R.~I.}\ \bibnamefont {Bewley}}, \bibinfo {author} {\bibfnamefont {D.~J.}\
			\bibnamefont {Voneshen}}, \bibinfo {author} {\bibfnamefont {M.~A.}\
			\bibnamefont {Green}}, \bibinfo {author} {\bibfnamefont {J.~A.}\ \bibnamefont
			{Rodriguez-Rivera}}, \bibinfo {author} {\bibfnamefont {P.}~\bibnamefont
			{Fouquet}}, \bibinfo {author} {\bibfnamefont {S.-W.}\ \bibnamefont {Cheong}},
		\bibinfo {author} {\bibfnamefont {J.~P.}\ \bibnamefont {Attfield}}, \bibinfo
		{author} {\bibfnamefont {R.~A.}\ \bibnamefont {Ewings}}, \ and\ \bibinfo
		{author} {\bibfnamefont {C.}~\bibnamefont {Stock}},\ }\bibfield  {title}
	{\enquote {\bibinfo {title} {Metastable antiphase boundary ordering in
				\textsc{C}a\textsc{F}e$_{2}$\textsc{O}$_{4}$},}\ }\href {\doibase
		10.1103/PhysRevB.104.104404} {\bibfield  {journal} {\bibinfo  {journal}
			{Phys. Rev. B}\ }\textbf {\bibinfo {volume} {104}},\ \bibinfo {pages}
		{104404} (\bibinfo {year} {2021})}\BibitemShut {NoStop}%
	\bibitem [{\citenamefont {Chan}\ \emph {et~al.}(2023)\citenamefont {Chan},
		\citenamefont {Lane}, \citenamefont {P\'asztorov\'a}, \citenamefont
		{Songvilay}, \citenamefont {Johnson}, \citenamefont {Downie}, \citenamefont
		{Bos}, \citenamefont {Rodriguez-Rivera}, \citenamefont {Cheong},
		\citenamefont {Ewings}, \citenamefont {Qureshi},\ and\ \citenamefont
		{Stock}}]{Chan23:107}%
	\BibitemOpen
	\bibfield  {author} {\bibinfo {author} {\bibfnamefont {E.}~\bibnamefont
			{Chan}}, \bibinfo {author} {\bibfnamefont {H.}~\bibnamefont {Lane}}, \bibinfo
		{author} {\bibfnamefont {J.}~\bibnamefont {P\'asztorov\'a}}, \bibinfo
		{author} {\bibfnamefont {M.}~\bibnamefont {Songvilay}}, \bibinfo {author}
		{\bibfnamefont {R.~D.}\ \bibnamefont {Johnson}}, \bibinfo {author}
		{\bibfnamefont {R.}~\bibnamefont {Downie}}, \bibinfo {author} {\bibfnamefont
			{J-W.~G.}\ \bibnamefont {Bos}}, \bibinfo {author} {\bibfnamefont {J.~A.}\
			\bibnamefont {Rodriguez-Rivera}}, \bibinfo {author} {\bibfnamefont {S.-W.}\
			\bibnamefont {Cheong}}, \bibinfo {author} {\bibfnamefont {R.~A.}\
			\bibnamefont {Ewings}}, \bibinfo {author} {\bibfnamefont {N.}~\bibnamefont
			{Qureshi}}, \ and\ \bibinfo {author} {\bibfnamefont {C.}~\bibnamefont
			{Stock}},\ }\bibfield  {title} {\enquote {\bibinfo {title} {Neutron
				scattering sum rules, symmetric exchanges, and helicoidal magnetism in
				\textsc{M}n\textsc{S}b$_{2}$\textsc{O}$_{6}$},}\ }\href {\doibase
		10.1103/PhysRevB.107.144420} {\bibfield  {journal} {\bibinfo  {journal}
			{Phys. Rev. B}\ }\textbf {\bibinfo {volume} {107}},\ \bibinfo {pages}
		{144420} (\bibinfo {year} {2023})}\BibitemShut {NoStop}%
	\bibitem [{\citenamefont {Brener}\ \emph {et~al.}(2024)\citenamefont {Brener},
		\citenamefont {Mallo}, \citenamefont {Lane}, \citenamefont
		{Rodriguez-Rivera}, \citenamefont {Schmalzl}, \citenamefont {Songvilay},
		\citenamefont {Guratinder}, \citenamefont {Petrovic},\ and\ \citenamefont
		{Stock}}]{Brener24:110}%
	\BibitemOpen
	\bibfield  {author} {\bibinfo {author} {\bibfnamefont {D.~J.}\ \bibnamefont
			{Brener}}, \bibinfo {author} {\bibfnamefont {I.~Rodriguez}\ \bibnamefont
			{Mallo}}, \bibinfo {author} {\bibfnamefont {H.}~\bibnamefont {Lane}},
		\bibinfo {author} {\bibfnamefont {J.~A.}\ \bibnamefont {Rodriguez-Rivera}},
		\bibinfo {author} {\bibfnamefont {K.}~\bibnamefont {Schmalzl}}, \bibinfo
		{author} {\bibfnamefont {M.}~\bibnamefont {Songvilay}}, \bibinfo {author}
		{\bibfnamefont {K.}~\bibnamefont {Guratinder}}, \bibinfo {author}
		{\bibfnamefont {C.}~\bibnamefont {Petrovic}}, \ and\ \bibinfo {author}
		{\bibfnamefont {C.}~\bibnamefont {Stock}},\ }\bibfield  {title} {\enquote
		{\bibinfo {title} {Anisotropic excitonic magnetism from discrete
				\textsc{C}$_{4}$ symmetry in \textsc{C}e\textsc{R}h\textsc{I}n$_{5}$},}\
	}\href {\doibase 10.1103/PhysRevB.110.064434} {\bibfield  {journal} {\bibinfo
			{journal} {Phys. Rev. B}\ }\textbf {\bibinfo {volume} {110}},\ \bibinfo
		{pages} {064434} (\bibinfo {year} {2024})}\BibitemShut {NoStop}%
	\bibitem [{\citenamefont {Buyers}\ \emph {et~al.}(1975)\citenamefont {Buyers},
		\citenamefont {Holden},\ and\ \citenamefont {Perreault}}]{Buyers75:11}%
	\BibitemOpen
	\bibfield  {author} {\bibinfo {author} {\bibfnamefont {W.~J.~L.}\
			\bibnamefont {Buyers}}, \bibinfo {author} {\bibfnamefont {T.~M.}\
			\bibnamefont {Holden}}, \ and\ \bibinfo {author} {\bibfnamefont
			{A.}~\bibnamefont {Perreault}},\ }\bibfield  {title} {\enquote {\bibinfo
			{title} {Temperature dependence of magnetic excitations in
				singlet-ground-state systems. ii. excited-state spin waves near the curie
				temperature in \textsc{P}r$_{3}$\textsc{T}l},}\ }\href {\doibase
		10.1103/PhysRevB.11.266} {\bibfield  {journal} {\bibinfo  {journal} {Phys.
				Rev. B}\ }\textbf {\bibinfo {volume} {11}},\ \bibinfo {pages} {266} (\bibinfo
		{year} {1975})}\BibitemShut {NoStop}%
	\bibitem [{\citenamefont {Telling}\ and\ \citenamefont
		{Andersen}(2005)}]{Telling05:7}%
	\BibitemOpen
	\bibfield  {author} {\bibinfo {author} {\bibfnamefont {M.~T.~F.}\
			\bibnamefont {Telling}}\ and\ \bibinfo {author} {\bibfnamefont {K.~H.}\
			\bibnamefont {Andersen}},\ }\bibfield  {title} {\enquote {\bibinfo {title}
			{Spectroscopic characteristics of the osris near-backscattering crystal
				analyser spectrometer on the isis pulsed neutron source},}\ }\href {\doibase
		10.1039/b413934h} {\bibfield  {journal} {\bibinfo  {journal} {Phys. Chem.
				Chem. Phys.}\ }\textbf {\bibinfo {volume} {7}},\ \bibinfo {pages} {1255}
		(\bibinfo {year} {2005})}\BibitemShut {NoStop}%
	\bibitem [{\citenamefont {Demmel}\ and\ \citenamefont
		{Pokhilchuk}(2014)}]{Demmel14:767}%
	\BibitemOpen
	\bibfield  {author} {\bibinfo {author} {\bibfnamefont {F.}~\bibnamefont
			{Demmel}}\ and\ \bibinfo {author} {\bibfnamefont {K.}~\bibnamefont
			{Pokhilchuk}},\ }\bibfield  {title} {\enquote {\bibinfo {title} {The
				resolution of the tof-backscattering spectrometer osiris: Monte carlo
				simulations and analytical calculations},}\ }\href {\doibase
		10.1016/j.nima.2014.09.019} {\bibfield  {journal} {\bibinfo  {journal}
			{NIM-A}\ }\textbf {\bibinfo {volume} {767}},\ \bibinfo {pages} {426}
		(\bibinfo {year} {2014})}\BibitemShut {NoStop}%
	\bibitem [{\citenamefont {Stock}\ \emph
		{et~al.}(2012{\natexlab{a}})\citenamefont {Stock}, \citenamefont {Broholm},
		\citenamefont {Zhao}, \citenamefont {Demmel}, \citenamefont {Kang},
		\citenamefont {Rule},\ and\ \citenamefont {Petrovic}}]{Stock12:109}%
	\BibitemOpen
	\bibfield  {author} {\bibinfo {author} {\bibfnamefont {C.}~\bibnamefont
			{Stock}}, \bibinfo {author} {\bibfnamefont {C.}~\bibnamefont {Broholm}},
		\bibinfo {author} {\bibfnamefont {Y.}~\bibnamefont {Zhao}}, \bibinfo {author}
		{\bibfnamefont {F.}~\bibnamefont {Demmel}}, \bibinfo {author} {\bibfnamefont
			{H.~J.}\ \bibnamefont {Kang}}, \bibinfo {author} {\bibfnamefont {K.~C.}\
			\bibnamefont {Rule}}, \ and\ \bibinfo {author} {\bibfnamefont
			{C.}~\bibnamefont {Petrovic}},\ }\bibfield  {title} {\enquote {\bibinfo
			{title} {Magnetic field splitting of the spin resonance in
				\textsc{C}e\textsc{C}o\textsc{I}n$_{5}$},}\ }\href {\doibase
		10.1103/PhysRevLett.109.167207} {\bibfield  {journal} {\bibinfo  {journal}
			{Phys. Rev. Lett.}\ }\textbf {\bibinfo {volume} {109}},\ \bibinfo {pages}
		{167207} (\bibinfo {year} {2012}{\natexlab{a}})}\BibitemShut {NoStop}%
	\bibitem [{\citenamefont {Stock}\ \emph
		{et~al.}(2012{\natexlab{b}})\citenamefont {Stock}, \citenamefont {Broholm},
		\citenamefont {Demmel}, \citenamefont {Van~Duijn}, \citenamefont {Taylor},
		\citenamefont {Kang}, \citenamefont {Hu},\ and\ \citenamefont
		{Petrovic}}]{Stock12:109_2}%
	\BibitemOpen
	\bibfield  {author} {\bibinfo {author} {\bibfnamefont {C.}~\bibnamefont
			{Stock}}, \bibinfo {author} {\bibfnamefont {C.}~\bibnamefont {Broholm}},
		\bibinfo {author} {\bibfnamefont {F.}~\bibnamefont {Demmel}}, \bibinfo
		{author} {\bibfnamefont {J.}~\bibnamefont {Van~Duijn}}, \bibinfo {author}
		{\bibfnamefont {J.~W.}\ \bibnamefont {Taylor}}, \bibinfo {author}
		{\bibfnamefont {H.~J.}\ \bibnamefont {Kang}}, \bibinfo {author}
		{\bibfnamefont {R.}~\bibnamefont {Hu}}, \ and\ \bibinfo {author}
		{\bibfnamefont {C.}~\bibnamefont {Petrovic}},\ }\bibfield  {title} {\enquote
		{\bibinfo {title} {From incommensurate correlations to mesoscopic spin
				resonance in \textsc{Y}b\textsc{R}h$_{2}$\textsc{S}i$_{2}$},}\ }\href
	{\doibase 10.1103/PhysRevLett.109.127201} {\bibfield  {journal} {\bibinfo
			{journal} {Phys. Rev. Lett.}\ }\textbf {\bibinfo {volume} {109}},\ \bibinfo
		{pages} {127201} (\bibinfo {year} {2012}{\natexlab{b}})}\BibitemShut
	{NoStop}%
	\bibitem [{\citenamefont {nd~E.~Ressouche}\ \emph {et~al.}(2004)\citenamefont
		{nd~E.~Ressouche}, \citenamefont {Brown}, \citenamefont {Henry},\ and\
		\citenamefont {Lelievre-Berna}}]{Kernavanois04:350}%
	\BibitemOpen
	\bibfield  {author} {\bibinfo {author} {\bibfnamefont {N.~Kernavanois}\
			\bibnamefont {nd~E.~Ressouche}}, \bibinfo {author} {\bibfnamefont {P.~J.}\
			\bibnamefont {Brown}}, \bibinfo {author} {\bibfnamefont {Y.~J.}\ \bibnamefont
			{Henry}}, \ and\ \bibinfo {author} {\bibfnamefont {E.}~\bibnamefont
			{Lelievre-Berna}},\ }\bibfield  {title} {\enquote {\bibinfo {title}
			{Magnetization distribution in paramagnetic nickel and cobalt oxides},}\
	}\href {\doibase 10.1016/j.physb.2004.03.066} {\bibfield  {journal} {\bibinfo
			{journal} {Physica B}\ }\textbf {\bibinfo {volume} {350}},\ \bibinfo {pages}
		{E265} (\bibinfo {year} {2004})}\BibitemShut {NoStop}%
	\bibitem [{\citenamefont {nd~E.~Ressouche}\ \emph {et~al.}(2003)\citenamefont
		{nd~E.~Ressouche}, \citenamefont {Brown}, \citenamefont {Henry},\ and\
		\citenamefont {Lelievre-Berna}}]{Kernavanois03:15}%
	\BibitemOpen
	\bibfield  {author} {\bibinfo {author} {\bibfnamefont {N.~Kernavanois}\
			\bibnamefont {nd~E.~Ressouche}}, \bibinfo {author} {\bibfnamefont {P.~J.}\
			\bibnamefont {Brown}}, \bibinfo {author} {\bibfnamefont {Y.~J.}\ \bibnamefont
			{Henry}}, \ and\ \bibinfo {author} {\bibfnamefont {E.}~\bibnamefont
			{Lelievre-Berna}},\ }\bibfield  {title} {\enquote {\bibinfo {title}
			{Magnetization distribution in paramagnetic \textsc{C}o\textsc{O}: a
				polarized neutron diffraction study},}\ }\href {\doibase
		10.1088/0953-8984/15/20/306} {\bibfield  {journal} {\bibinfo  {journal} {J.
				Phys.: Condens. Matter}\ }\textbf {\bibinfo {volume} {15}},\ \bibinfo {pages}
		{3433} (\bibinfo {year} {2003})}\BibitemShut {NoStop}%
	\bibitem [{\citenamefont {Zhitomirsky}\ and\ \citenamefont
		{Chernyshev}(2013)}]{Zhitomirsky13:85}%
	\BibitemOpen
	\bibfield  {author} {\bibinfo {author} {\bibfnamefont {M.~E.}\ \bibnamefont
			{Zhitomirsky}}\ and\ \bibinfo {author} {\bibfnamefont {A.~L.}\ \bibnamefont
			{Chernyshev}},\ }\bibfield  {title} {\enquote {\bibinfo {title} {Colloquium:
				Spontaneous magnon decays},}\ }\href
	{https://link.aps.org/doi/10.1103/RevModPhys.85.219} {\bibfield  {journal}
		{\bibinfo  {journal} {Rev. Mod. Phys.}\ }\textbf {\bibinfo {volume} {85}},\
		\bibinfo {pages} {219} (\bibinfo {year} {2013})}\BibitemShut {NoStop}%
	\bibitem [{\citenamefont {Masuda}\ \emph {et~al.}(2006)\citenamefont {Masuda},
		\citenamefont {Zheludev}, \citenamefont {Manaka}, \citenamefont {Regnault},
		\citenamefont {Chung},\ and\ \citenamefont {Qiu}}]{Masuda06:96}%
	\BibitemOpen
	\bibfield  {author} {\bibinfo {author} {\bibfnamefont {T.}~\bibnamefont
			{Masuda}}, \bibinfo {author} {\bibfnamefont {A.}~\bibnamefont {Zheludev}},
		\bibinfo {author} {\bibfnamefont {H.}~\bibnamefont {Manaka}}, \bibinfo
		{author} {\bibfnamefont {L.-P.}\ \bibnamefont {Regnault}}, \bibinfo {author}
		{\bibfnamefont {J.-H.}\ \bibnamefont {Chung}}, \ and\ \bibinfo {author}
		{\bibfnamefont {Y.}~\bibnamefont {Qiu}},\ }\bibfield  {title} {\enquote
		{\bibinfo {title} {Dynamics of composite haldane spin chains in
				\textsc{IPA}-\textsc{C}u\textsc{C}l$_{3}$},}\ }\href {\doibase
		10.1103/PhysRevLett.96.047210} {\bibfield  {journal} {\bibinfo  {journal}
			{Phys. Rev. Lett.}\ }\textbf {\bibinfo {volume} {96}},\ \bibinfo {pages}
		{047210} (\bibinfo {year} {2006})}\BibitemShut {NoStop}%
	\bibitem [{\citenamefont {Stanley}(1971)}]{Stanley:book}%
	\BibitemOpen
	\bibfield  {author} {\bibinfo {author} {\bibfnamefont {H.~E.}\ \bibnamefont
			{Stanley}},\ }\href@noop {} {\emph {\bibinfo {title} {Introduction to Phase
				Transitions and Critical Phenomena}}}\ (\bibinfo  {publisher} {Oxford
		University Press},\ \bibinfo {address} {Oxford},\ \bibinfo {year}
	{1971})\BibitemShut {NoStop}%
	\bibitem [{\citenamefont {Wehner}\ and\ \citenamefont
		{Steigmeier}(1975)}]{Wehner75:36}%
	\BibitemOpen
	\bibfield  {author} {\bibinfo {author} {\bibfnamefont {R.~K.}\ \bibnamefont
			{Wehner}}\ and\ \bibinfo {author} {\bibfnamefont {E.~F.}\ \bibnamefont
			{Steigmeier}},\ }\bibfield  {title} {\enquote {\bibinfo {title} {Coupled
				lattice modes in light scattering},}\ }\href@noop {} {\bibfield  {journal}
		{\bibinfo  {journal} {RCA Review}\ }\textbf {\bibinfo {volume} {36}},\
		\bibinfo {pages} {70} (\bibinfo {year} {1975})}\BibitemShut {NoStop}%
	\bibitem [{\citenamefont {Magarino}\ \emph {et~al.}(1977)\citenamefont
		{Magarino}, \citenamefont {Tuchendler}, \citenamefont {Fert},\ and\
		\citenamefont {Gelard}}]{Magarino77:23}%
	\BibitemOpen
	\bibfield  {author} {\bibinfo {author} {\bibfnamefont {J.}~\bibnamefont
			{Magarino}}, \bibinfo {author} {\bibfnamefont {J.}~\bibnamefont
			{Tuchendler}}, \bibinfo {author} {\bibfnamefont {A.~R.}\ \bibnamefont
			{Fert}}, \ and\ \bibinfo {author} {\bibfnamefont {J.}~\bibnamefont
			{Gelard}},\ }\bibfield  {title} {\enquote {\bibinfo {title} {Field dependence
				of uniform magnon energies in lamellar \textsc{C}o\textsc{C}l$_{2}$ and
				\textsc{C}o\textsc{B}r$_{2}$ by \textsc{AFMR} experiments},}\ }\href
	{\doibase 10.1016/0038-1098(77)90103-X} {\bibfield  {journal} {\bibinfo
			{journal} {Solid State Commun.}\ }\textbf {\bibinfo {volume} {23}},\ \bibinfo
		{pages} {175} (\bibinfo {year} {1977})}\BibitemShut {NoStop}%
	\bibitem [{\citenamefont {R\"uegg}\ \emph {et~al.}(2008)\citenamefont
		{R\"uegg}, \citenamefont {Normand}, \citenamefont {Matsumoto}, \citenamefont
		{Furrer}, \citenamefont {McMorrow}, \citenamefont {Kr\"amer}, \citenamefont
		{G\"udel}, \citenamefont {Gvasaliya}, \citenamefont {Mutka},\ and\
		\citenamefont {Boehm}}]{Ruegg08:100}%
	\BibitemOpen
	\bibfield  {author} {\bibinfo {author} {\bibfnamefont {Ch.}\ \bibnamefont
			{R\"uegg}}, \bibinfo {author} {\bibfnamefont {B.}~\bibnamefont {Normand}},
		\bibinfo {author} {\bibfnamefont {M.}~\bibnamefont {Matsumoto}}, \bibinfo
		{author} {\bibfnamefont {A.}~\bibnamefont {Furrer}}, \bibinfo {author}
		{\bibfnamefont {D.~F.}\ \bibnamefont {McMorrow}}, \bibinfo {author}
		{\bibfnamefont {K.~W.}\ \bibnamefont {Kr\"amer}}, \bibinfo {author}
		{\bibfnamefont {H.~U.}\ \bibnamefont {G\"udel}}, \bibinfo {author}
		{\bibfnamefont {S.~N.}\ \bibnamefont {Gvasaliya}}, \bibinfo {author}
		{\bibfnamefont {H.}~\bibnamefont {Mutka}}, \ and\ \bibinfo {author}
		{\bibfnamefont {M.}~\bibnamefont {Boehm}},\ }\bibfield  {title} {\enquote
		{\bibinfo {title} {Quantum magnets under pressure: Controlling elementary
				excitations in \textsc{T}l\textsc{C}u\textsc{C}l$_{3}$},}\ }\href {\doibase
		10.1103/PhysRevLett.100.205701} {\bibfield  {journal} {\bibinfo  {journal}
			{Phys. Rev. Lett.}\ }\textbf {\bibinfo {volume} {100}},\ \bibinfo {pages}
		{205701} (\bibinfo {year} {2008})}\BibitemShut {NoStop}%
	\bibitem [{\citenamefont {Coldea}\ \emph {et~al.}(2010)\citenamefont {Coldea},
		\citenamefont {Tennant}, \citenamefont {Wheeler}, \citenamefont {Wawrzynska},
		\citenamefont {Prabhakaran}, \citenamefont {Telling}, \citenamefont
		{Habicht}, \citenamefont {Smeibidl},\ and\ \citenamefont
		{Kiefer}}]{Coldea10:327}%
	\BibitemOpen
	\bibfield  {author} {\bibinfo {author} {\bibfnamefont {R.}~\bibnamefont
			{Coldea}}, \bibinfo {author} {\bibfnamefont {D.~A.}\ \bibnamefont {Tennant}},
		\bibinfo {author} {\bibfnamefont {E.~M.}\ \bibnamefont {Wheeler}}, \bibinfo
		{author} {\bibfnamefont {E.}~\bibnamefont {Wawrzynska}}, \bibinfo {author}
		{\bibfnamefont {D.}~\bibnamefont {Prabhakaran}}, \bibinfo {author}
		{\bibfnamefont {M.}~\bibnamefont {Telling}}, \bibinfo {author} {\bibfnamefont
			{K.}~\bibnamefont {Habicht}}, \bibinfo {author} {\bibfnamefont
			{P.}~\bibnamefont {Smeibidl}}, \ and\ \bibinfo {author} {\bibfnamefont
			{K.}~\bibnamefont {Kiefer}},\ }\bibfield  {title} {\enquote {\bibinfo {title}
			{Quantum criticality in an ising chain: Experimental evidence for emergent
				\textsc{E}$_{8}$ symmetry},}\ }\href {\doibase 10.1126/science.1180085}
	{\bibfield  {journal} {\bibinfo  {journal} {Science}\ }\textbf {\bibinfo
			{volume} {327}},\ \bibinfo {pages} {177} (\bibinfo {year}
		{2010})}\BibitemShut {NoStop}%
\end{thebibliography}

%

\end{document}


\title{Supplementary Information for ``Zeeman split Kramers doublets in spin-supersolid candidate Na$_{2}$BaCo(PO$_{4}$)$_{2}$"}

\author{T. I. Popescu}
\affiliation{School of Physics and Astronomy, University of Edinburgh, EH9 3JZ, United Kingdom}
\author{N. Gora}
\affiliation{School of Physics and Astronomy, University of Edinburgh, EH9 3JZ, United Kingdom}
\author{F. Demmel}
\affiliation{ISIS Facility, Rutherford Appleton Labs, Chilton, Didcot, OX11 0QX}
\author{Z. Xu}
\affiliation{NIST Center for Neutron Research, 100 Bureau Drive, Gaithersburg, Maryland 20899, USA}
\author{R. Zhong}
\affiliation{Department of Chemistry, Princeton University, Princeton NJ 08544, USA}
\author{T. J. Williams}
\affiliation{ISIS Pulsed Neutron and Muon Source, STFC Rutherford Appleton Laboratory, Harwell Campus, Didcot, Oxon OX11 0QX, United Kingdom}
\author{R. J. Cava}
\affiliation{Department of Chemistry, Princeton University, Princeton NJ 08544, USA}
\author{G. Xu}
\affiliation{NIST Center for Neutron Research, 100 Bureau Drive, Gaithersburg, Maryland 20899, USA}
\author{C. Stock}
\affiliation{School of Physics and Astronomy, University of Edinburgh, EH9 3JZ, United Kingdom}

\date{\today}

\begin{abstract}

In this supplementary information we provide more details on the theoretical formulation of the Green's function response applied to neutron scattering and also provide more details on the experiment and calculations to understand the magnetic phase present at intermediate fields.  This field range was denoted as the $|\uparrow \uparrow \downarrow \rangle$ phase in the main text.

\end{abstract}

\pacs{}
\maketitle

\section{Theoretical Formulation of the Green's Function Response}

We begin by discussing the Green's function formalism employed in this study. We recall that the neutron magnetic cross section is proportional to the magnetic dynamic structure factor, $S(\boldsymbol{Q}, \omega)$, given by 

\begin{equation} \label{eq:1}
S(\boldsymbol{Q}, \omega) = g_L^2 f^2(\boldsymbol{Q}) \sum_{\alpha\beta} (\delta_{\alpha\beta} - Q_\alpha Q_\beta) S^{\alpha\beta}(\boldsymbol{Q}, \omega),
\end{equation}

\begin{table}[h]
	\begin{ruledtabular} \label{Table_notation}
		\begin{tabular}{cc}
			Index & Description \\ 
			\hline 
			$i$, $j$, .. & unit cell position indices  \\ 

   $\alpha$, $\beta$, $\mu$, $\nu$,... & Cartesian coordinates $(x,y,z)$  \\ 
			$A, B, C,...$ & Quantities in lab frame \\
			$\tilde{A}, \tilde{B}, \tilde{C},...$ & Quantities in local rotating frame \\ 
			$\mathcal{J}$ & Symmetric (scalar) Heisenberg exchange \\ 
		\end{tabular}
	\end{ruledtabular}
 \caption{\label{Table:indices} Summary of labeling convention for indices and notation.}
\end{table} 

\noindent where $g_{L}$ is the Lande $g$ factor, $f(\boldsymbol{Q})$ is the magnetic form factor and $S^{\alpha\beta}(\boldsymbol{Q}, \omega)$ is the dynamic spin structure constant with the sum being taken over the Cartesian coordinates $\alpha, \beta$. For $3d^7$ Co$^{2+}$, angular momentum is quenched ($\langle L \rangle$ = 0), therefore it can be assumed that the spin operators provide the dominant contribution to the scattering cross section, as opposed to the orbital angular momentum.

We can define the equation of motion of our response function as, 
\begin{equation} \label{eq:2}
\omega G(\mathit{A}, \mathit{B}, \omega) = \langle [\mathit{A}, \mathit{B}] \rangle + G([\mathit{A}, \mathit{\mathcal{H}}], \mathit{B}, \omega),
\end{equation}
where $\mathit{A}$ and $\mathit{B}$ denote generic spin operators. To solve equation (\ref{eq:2}), we require both an expression for our Hamiltonian $\mathit{\mathcal{H}}$ but also for its commutator with the spin operators. 
\subsection{Total Magnetic Hamiltonian, $\mathit{\mathcal{H}}$}
We begin by defining our total Hamiltonian as
\begin{equation}
\mathit{\mathcal{H}} = \mathit{\mathcal{H}}_{\text{CF}} + \sum_{ij} \mathcal{J}(ij) \mathit{\boldsymbol{S}}(i) \cdot \mathit{\boldsymbol{S}}(j),
\end{equation}
with contributions from the crystal field (CF) and the coupling between the Co$^{2+}$ on sites $i,j$ where $\mathcal{J}(ij)$ is the exchange coupling constant. 
The total magnetic Hamiltonian for a lattice can be represented as the sum of the crystal field (CF) contributions along with the coupling between Co$^{2+}$ ions on sites $i$ and $j$, given by 

\begin{equation}
\mathcal{H} = \mathcal{H}_{\text{CF}} + \sum_{ij} \mathcal{J}(ij) \mathit{\boldsymbol{S}}(i) \cdot \mathit{\boldsymbol{S}}(j),
\end{equation}

\noindent where $J(ij)$ is the exchange coupling constant. The molecular field Hamiltonian can then be defined as 

\begin{equation}
\mathcal{H}_{\text{MF}}(i) = \sum_i h_{\text{MF}}(i) \mathit{S}_z(i),
\end{equation}

\noindent where $h_{\text{MF}}(i)$ is given by

\begin{equation}
h_{\text{MF}}(i) = 2 \sum_{j} \mathcal{J}(ij) \langle \mathit{S}_z(j) \rangle,
\end{equation}

\noindent where the average spin due to thermal fluctuation is given by $\langle \mathit{S}_z(j) \rangle = \sum_n S_{znn} \ f_{n}$ with the Boltzmann population factor being defined as\[f_{n} = \frac{\exp(- \omega_{n}/k_bT)}{\sum_m \exp(- \omega_{m}/k_bT)}\] and $S_{znn} = \langle n | \mathit{S}_{z} | n \rangle$.

By expanding the total Hamiltonian, and adding and subtracting the molecular field from the expression, the total Hamiltonian can be written as both a single-ion component ($\mathit{\mathcal{H}_1}$) and an interion part ($\mathit{\mathcal{H}_2}$), each given by,

\begin{equation}
\mathit{\mathcal{H}_1} = \sum_i \mathit{\mathcal{H}}_{\text{CF}}(i) + \sum_i \mathit{S}_z(i) \left( 2 \sum_{j} \mathcal{J}(ij) \langle \mathit{S}_z(j) \rangle \right),
\end{equation}

and

\begin{align}
\mathit{\mathcal{H}}_2 &= \sum_{ij} \mathcal{J}(ij) \mathit{S}_z(i) [\mathit{S}_z(j) - 2 \langle \mathit{S}_z(j) \rangle] \notag \\
&\quad + \frac{1}{2} \sum_{ij} \mathcal{J}(ij) [S_+(i) S_-(j) + S_-(i) S_+(j)],
\end{align}

To derive the neutron response, ${\mathcal{H}_1}$ will be diagonalised using the ladder operators $C^{\dagger}(i)$ and $C(i)$ (which satisfy $[C^{\dagger}_n (i),C_m (j)] = \delta_{ij}\delta_{nm}$), to give

\begin{equation}
\mathcal{H}_1 = \sum_n \sum_i \omega_n C_n^\dagger(i) C_n(i),
\end{equation}

\noindent where $\omega_n$ are the energy eigenvalues. By using mean field theory on the interion term ($\mathcal{H}_2$), the total Hamiltonian can be used in the equation of motion of the response function. Firstly, however, the eigenstates of the single-ion Hamiltonian must be derived. 
\subsection{Eigenstates of single-ion Hamiltonian}

We begin with the derivation of the single ion Hamiltonian, expanding the single-ion component as

\begin{equation}
\mathcal{H}_1 = \mathcal{H}_{\text{CF}} + \mathcal{H}_{\text{MF}} 
= (\mathcal{H}_{\text{CEF}} + \mathcal{H}_{\text{SO}} + \mathcal{H}_{\text{dis}}) +\mathcal{H}_{\text{MF}},
\end{equation}

\noindent where $\mathcal{H}_{\text{CEF}}$ are contributions due to the crystal electric field of the structure, $\mathcal{H}_{\text{SO}}$ are spin-orbit contributions,  $\mathcal{H}_{\text{dis}}$ are contributions from structural distortion and $\mathcal{H}_{\text{MF}}$ is the mean molecular field. 
The \textit{Crystalline Electric Field} contribution corresponding to the hexagonal structure, specifically the point group symmetry of $D_{6h}$ of a Co$^{2+}$ compound, of the free-ion surrounded by the ligand O simplify the general equation for CEF, using Steven's operators ($\mathcal{O}_k^q$), from 

\begin{equation}
\mathcal{H}_{\text{CEF}} = \sum_{k,q} B_k^q \mathcal{O}_k^q,
\end{equation}

\noindent to 

\begin{equation}
\mathcal{H}_{\text{CEF}} = B_2^0 \mathcal{O}_2^0 + B_4^0 \mathcal{O}_4^0.
\end{equation}

\noindent If the crystal-field energy splitting for the hexagonally coordinated Co$^{2+}$ is assumed to be far less than the energy differences between the free-ion terms, then the crystal-field Hamiltonian can be treated as a perturbation to the free-ion bases states, which are in turn defined by Hund's rules. Using these rules (i.e. accounting for electron-electron interactions and Pauli's exclusion principle), the corresponding term symbol for the ground state is $^4T$.

Since spin-orbit coupling is significantly weaker that the crystal electric field, ${L}^2$, ${S}^2$, ${L}_z$ and ${S}_z$ remain as good quantum numbers under the Russel-Saunders L-S coupling scheme since the ground state triplet in Co$^{2+}$ spans the subspace $|l =1, m_l\rangle$ rather than $|L =3, m_l\rangle$. Hence all consequent Hamiltonians can be projected onto this space.

In this, we relegate all other parts of the Hamiltonian to simply perturbation additions to the $^4T$ free-ion ground state. To do this in practice, we diagonalise the single-ion term excluding $\mathcal{H}_{\text{CEF}}$, then rotate all spin terms of S-O into this basis, knowing $|l =1, m_l\rangle$  already. 

Further, the molecular field emerges from the effect of the ligands on the ordering of Co$^{2+}$. $\mathcal{H}_{\text{MF}}$ behaves as a Zeeman term, splitting the terms nearly degenerate $j_\text{eff}$ levels after the effects of spin-orbit coupling. This Hamiltonian can be defined by considering the dominant nearest neighbor as:
\begin{equation}
H_{\text{MF}} = \sum_i h_{\text{MF}}(i) S_z = 2 z_1 \mathcal{J}_1 \langle S_z \rangle S_z,
\end{equation}
where \( z_1 \) is the number of nearest neighbors and \(\mathcal{ J}_1\) 
is the nearest neighbor exchange constant.

\subsection{Mean Field Theory for Spin-Orbit Excitations}

As previously mentioned, to model our neutron response function we require an understanding of both the Hamiltonian, together with how it commutes with the spin operators. Since we wish to have both the single-ion and interion component of the Hamiltonian be in the same basis, we can rotate $\mathcal{H}_{2}$ into our $\mathcal{H}_{1}$ basis. This is possible because $\mathcal{H}_{2}$ is based on the components of the spin operators $\mathbf{S}$. This method is particularly convenient as the single-ion Hamiltonian $\mathcal{H}_{1}$, is already diagonalized. To perform this rotation of the spin operators, the set of ladder operators previously defined above can be employed to write this rotation as
\begin{align}
S_{(\pm,z)}=\sum_{mn}S_{(\pm,z)mn}C^{\dagger}_mC_n.
\end{align}
To begin obtaining an expression for our Green's function we can define the interlevel susceptibility similar to previous studies \cite{Sarte19:100,Collins97:75}  as,  
\begin{align}
G^{\alpha\beta}(ij,\omega)=\sum_{mn}S_{\alpha mn}\hat{G}^{\beta}(mn,ij,\omega),
\end{align}
where it is possible to solve for $\hat{G}^{\beta}(A,B,\omega)$ using the Heisenberg EOM where $A=C^{\dagger}_mC_n$ and $B=S_{\beta}$. This follows from our definition of $G^{\alpha\beta}$ which when written in terms of $\hat{G}^{\beta}$ yields

\begin{align}
\sum_{mn}S_{\alpha mn}\hat{G}^{\beta}(mn,ij,\omega)=\int dt e^{i\omega t}\left(-i\Theta(t)\langle[S_{\alpha},S_{\beta}]\rangle \right).
\end{align}

When writing $S_{\alpha}$ in our $\mathcal{H}_{1}$ basis we can simplify the previous expression to

\begin{align}
\sum_{mn}S_{\alpha mn}\hat{G}^{\beta}(mn,ij,\omega)=...\\
\int dt e^{i\omega t}\left(-i\Theta(t)\langle[\sum_{mn}S_{\alpha mn}C^{\dagger}_mC_n,S_{\beta}]\rangle \right),
\end{align}

\noindent and since $S_{\alpha mn}$ is a $C$-number it is possible to obtain after Fourier transforming the left-hand side 

\begin{align}
\hat{G}^{\beta}(mn,ij,t)=-i \Theta(t)\langle[C^\dagger_mC_n,S_\beta]\rangle,
\end{align}
thus confirming our choices for $A$ and $B$.
\subsection{Derivation of Commutators}
When obtaining an equation for $\hat{G}$, there will be three seperate terms: diagonal, transverse and longitudinal. The commutator of $C^{\dagger}_mC_n$ with the diagonal terms of $\mathcal{H}_1$ yields diagonal terms of the form
\begin{align}
\sum_k\sum_{r}[C^{\dagger}_m(i)C_n(j),C^{\dagger}_r(k)C_r(k)]\omega_r,
\end{align}
which when combined with the condition that our ladder operators must satisfy
\begin{align}
[C^{\dagger}_{m}C_n,C^{\dagger}_rC_s]=\delta_{nr}C^{\dagger}_mC_s-\delta_{sm}C^{\dagger}_rC_n,
\end{align}
gives a single non-zero term
\begin{align}
\delta_{ij}(\omega_n-\omega_m)C^{\dagger}_m(i)C_n(j),
\end{align}
having inserted the $\delta_{ij}$ to account for the commutator vanishing when our ladder operators act on different spaces. The commutator of $C^{\dagger}_mC_n$ with the transverse terms in our Hamiltonian yielded a similar equation of the form 
\begin{align}
\sum_{kl}\mathcal{J}_{kl}[C^{\dagger}_m(i)C_{n}(i),S_{+}(k)S_{-}(l)].
\end{align}
Proceeding with our expressions for the spin operators in the $\mathcal{H}_{1}$ basis we obtain
\begin{align}
&\sum_{kl}\mathcal{J}_{kl}[C^{\dagger}_m(i)C_{n}(i),\notag \\
&\quad \sum_{pq}S_{+pq}(k)C^{\dagger}_p(k)C_{q}(k)\sum_{rs}S_{-rs}(l)C^{\dagger}_r(l)C_{s}(l)],
\end{align}
which can be simplified to
\begin{align}
&\sum_{kl}\sum_{pq}\sum_{rs}\mathcal{J}_{kl}S_{+pq}(k)S_{-rs}(l)\notag \\
&\quad[C^{\dagger}_m(i)C_{n}(i),C^{\dagger}_p(k)C_{q}(k)C^{\dagger}_r(l)C_{s}(l)].
\end{align}
Applying random phase decoupling 
\begin{align}
&C^{\dagger}_p(k)C_{q}(k)C^{\dagger}_r(l)C_{s}(l)\simeq \notag \\
&\quad f_{p}(k)\delta_{pq}C^{\dagger}_r(l)C_s(l)+f_r(l)\delta_{rs}C^{\dagger}_{p}(k)C_{q}(k)
\end{align}

\noindent we can re-express our transverse term as 

\begin{align}\label{EQN21}
&\sum_{kl}\sum_{pq}\sum_{rs}\mathcal{J}_{kl}S_{+pq}(k)S_{-rs}(l)(f_{p}(k)\delta_{pq}\notag \\
&\quad[C^{\dagger}_m(i)C_{n}(i),C^{\dagger}_r(l)C_s(l)]+\notag \\
&\quad f_r(l)\delta_{rs}[C^{\dagger}_m(i)C_{n}(i),C^{\dagger}_{p}(k)C_{q}(k)]).
\end{align}
Using one of the properties of our ladder operators
\begin{align}
\sum_{rs}[C^{\dagger}_mC_{n},C^{\dagger}_rC_s]S_{\alpha r s}=\sum_s(C^{\dagger}_mC_sS_{\alpha n s}-C^{\dagger}_sC_nS_{\alpha s m}),
\end{align}
it is possible to re-express (\ref{EQN21}) as 
\begin{align}
&\sum_{l}\mathcal{J}_{il}S_{+nm}(f_m-f_n)\sum_{-pq} \\
&\quad C^{\dagger}_qC_{p}+S_{-nm}(f_m-f_n)\sum_{pq}S_{+qp}C^{\dagger}_{q}C_{p}.
\end{align}
Lastly, we compute the longitudinal commutator given by
\begin{align}
\sum_{lk}\mathcal{J}_{lk}[C^{\dagger}_mC_{n},S_z(k)\left\{S_z(l)-2\langle S_{z}(l)\rangle \right\}],
\end{align}
which when simplified using our ladder operator properties becomes
\begin{align}
2\sum_{l}\mathcal{J}_{il}S_{znm}(f_m-f_n)\sum_{pq}S_{zpq}C^{\dagger}_q(l)C_p(l)
.
\end{align}
The final commutator required to be able to solve the Heisenberg equation of motion is $\langle[A,B]\rangle$. This commutator conveniently simplifies to 
\begin{align}
\langle[A,B]\rangle=(f_m(i)-f_n(i))S_{\beta nm}(j)\delta_{ij}.
\end{align}
\subsection{Obtaining the Green's Function}
The single-site dynamical susceptibility, $g^{\alpha \beta}(\Vec{q},\omega)$ is the solution to the interlevel susceptibility for $\mathcal{J}_{il}=0$ which becomes
\begin{align}
&\omega \hat{G}(mn,ij,\omega)=(f_m-f_n)S_{\beta mn}\delta_{ij}+\notag\\
&\quad(\omega_n-\omega_m)\hat{G}(mn,ij,\omega).
\end{align}

After performing a Fourier transform, and multiplying with $S_{\alpha mn}$ we obtain 

\begin{align}
g^{\alpha \beta}(\omega)=\sum_{mn}\frac{S_{\alpha mn}S_{\beta nm}(f_m-f_n)}{\omega-\omega_{n}+\omega_{m}+i\chi}.
\end{align}

\noindent The additional parameter $i\chi$ moves the singularities (poles) off the real axis to ensure the response function is analytic.  The parameter $\chi$ can be physically interpreted as a resolution width as it broadens the response in energy.  We have fixed this parameter to the resolution of OSIRIS discussed in the main text.

Applying the same method for the whole Green's function yields
\begin{align}
&G^{\alpha \beta}(\textbf{q},\omega)=g^{\alpha \beta}(\omega)+g^{\alpha +}(\omega)\mathcal{J}(\textbf{q})G^{-\beta}(\textbf{q},\omega)+\notag\\
&g^{\alpha-}(\textbf{q},\omega)\mathcal{J}(\textbf{q})G^{+\beta}(\textbf{q},\omega)+\notag\\
&2g^{\alpha z}(\omega)\mathcal{J}(\textbf{q})G^{z\beta}(\textbf{q},\omega).
\end{align}
Where $J(\vec{q})$ is the Fourier transform of the exchange interaction. Due to the high symmetry of the lattices we will examine, there are only three non-zero components of the single-ion susceptibility, these are: $g^{+-}$,$g^{-+}$ and $g^{zz}$. Thus the solution for the spin wave is then
\begin{align}
G^{+-}(\mathbf{q},\omega)=\frac{g^{+-}(\omega)}{1-\mathcal{J}(\mathbf{q})g^{+-}(\omega)},
\end{align}
for the transverse modes and 
\begin{align}
G^{zz}(\mathbf{q},\omega)=\frac{g^{zz}(\omega)}{1-2\mathcal{J}(\mathbf{q})g^{zz}(\omega)},
\end{align}
for the longitudinal mode. 

\subsection{Rotating Frame Formalism}
The material under investigation can be considered to have a non-collinear magnetic structure as a result of each unit cell having a different mean field Hamiltonian. This can be overcome by transforming to a coordinate frame that rotates with the magnetic structure of the sample. We choose this rotating frame such that the magnetic moment is orientated along the $\tilde{z}$ axis. The spin vector in the lab frame is related to the spin vector in the rotating frame by,

\begin{align}
S_{i\gamma}=R_{i\gamma}\tilde{S}_{i\gamma}
\end{align}

with the tilde denoting the rotating frame. The rotation can be broken into two components, rotating the spins in the unit cell onto a common basis for the unit cell and a rotation of the unit cell onto a common rotating frame basis. In order to relate the rotation of each unit cell $R_{i}$ to the magnetic ordering wavevector $\mathbf{Q}$ and spin rotation plane $\mathbf{n}$, we employ the Rodrigues formula

\begin{align}
R_{i}=e^{i\mathbf{Q}\cdot\mathbf{r}_{i}}T+e^{-i\mathbf{Q}\cdot\mathbf{r}_{i}}T^*+\mathbf{nn}^{T},
\end{align}

\begin{align}
T=\frac{1}{2}(\mathbf{1}-\mathbf{nn}^{T}-i[\mathbf{n}]_{\times}).
\end{align}
Thus, by taking a Fourier transform of the Green's function using the rotating frame formalism we obtain
\begin{align}
{{\tilde{G}}}(\mathbf{q},\omega)={{g}}(\omega)+{{g}}(\omega){{\tilde{\mathcal{J}}}}(\mathbf{q}){{\tilde{G}}}(\mathbf{q},\omega),
\end{align}
where we define $\mathcal{J}$ as
\begin{align}
{{\mathcal{J}}}(\mathbf{q})=\frac{1}{N}\sum_{mn}{{\mathcal{J}_{mn}}}e^{-i\mathbf{q}(\mathbf{r}_m-\mathbf{r}_n)},
\end{align}
where N is the number of nearest neighbors.
Writing $\mathcal{J}$ in the rotating frame formalism we get
\begin{align}
{{\tilde{\mathcal{J}}}}(\mathbf{q})={{\mathcal{J}}}(\mathbf{q}+\mathbf{Q})\Phi+{{\mathcal{J}}}(\mathbf{q}-\mathbf{Q})\Phi^{*}+{{\mathcal{J}}}(\mathbf{q})\mathbf{n}\mathbf{n}^{T}.
\end{align}
Using the Fourier transform of the exchange interaction $J(\mathbf{Q})$, we obtain the non-zero components of $G(\mathbf{Q},\omega)$ as,

\begin{align}
G(\mathbf{Q},\omega)= G^{+-}(\mathbf{Q},\omega)+G^{-+}(\mathbf{Q},\omega)+G^{zz}(\mathbf{Q},\omega).
\end{align}

To account for the observed strong magnon decay, we employ the imaginary-part self-consistent Born approximation (iSCBA). Through the use of the iSCBA, our ansatz for the decay corrected response function takes the form, 
 
\begin{equation}
    \mathcal{G}^{-1}(\mathbf{Q},\omega)=G^{-1}(\mathbf{Q},\omega)+i\kappa,
\end{equation}
where $\mathcal{G}(\mathbf{Q},\omega)$ is the magnon decay corrected Green's function and $K$ is the magnon decay rate \cite{Zhitomirsky13:85}. Taking the self-consistent form of $K$ to be an analytic function of $\mathbf{Q}$ and knowing it must be even in $\mathbf{Q}$ from Bloch's theorem, we can re-express it as a power series. Keeping only the first $\mathbf{Q}^{2}$ term in the expansion yields

\begin{equation}
    \kappa=\xi\mathbf{Q}^2,
\end{equation}
where $\xi$ is some variational parameter.

Since the imaginary component of the total response function is proportional to the dynamical structure factor in the T$\rightarrow$ 0 K limit, equation (\ref{eq:1}) can be reduced to

\begin{equation}
    S(\mathbf{Q},\omega)\underset{\sim}{\propto}-f(\mathbf{Q})\mathcal{I}\mathcal{G}(\mathbf{Q},\omega),
\end{equation}

which is our theoretical response.
\subsection{Multi-Ion Excitations}
The Hamiltonian for an ion cluster for an isotropic configuration was taken to be
\begin{align}
    &H =\sum_{i,j} \mathcal{J}_1 \mathbf{S}(i)\cdot\mathbf{S}(j)+ \sum_{i}\mathcal{H}_1(i).
\end{align}
The determined pair spectra for the isotropic configuration for the two field measurements in the $|\hspace{-2pt}\uparrow\uparrow\downarrow\rangle$ phase are presented in \ref{figure_5}[(a) (b)]. The decrease in the gap of the two levels originates from the molecular field experienced by the anti-aligned spin in the $\uparrow \downarrow$ pair undergoing the excitation. As the field is decreased towards the liquid like ground state, the parameter for molecular field of this anti-aligned spin sharply decreases from the saturation value of 0.13 meV to 0.07 meV. This decrease in molecular field can be interpreted as increasing the number of anti-aligned spins with the external field. The resulting decrease in entropy to accommodate for the energetically costly spin flips results in an increase in temperature making NBCP an inefficient adiabatic coolant in the intermediate $|\hspace{-2pt}\uparrow\uparrow\downarrow\rangle$ phase.

\renewcommand{\thefigure}{S1}
\begin{figure}[t] 
	\includegraphics[ width=1\linewidth]{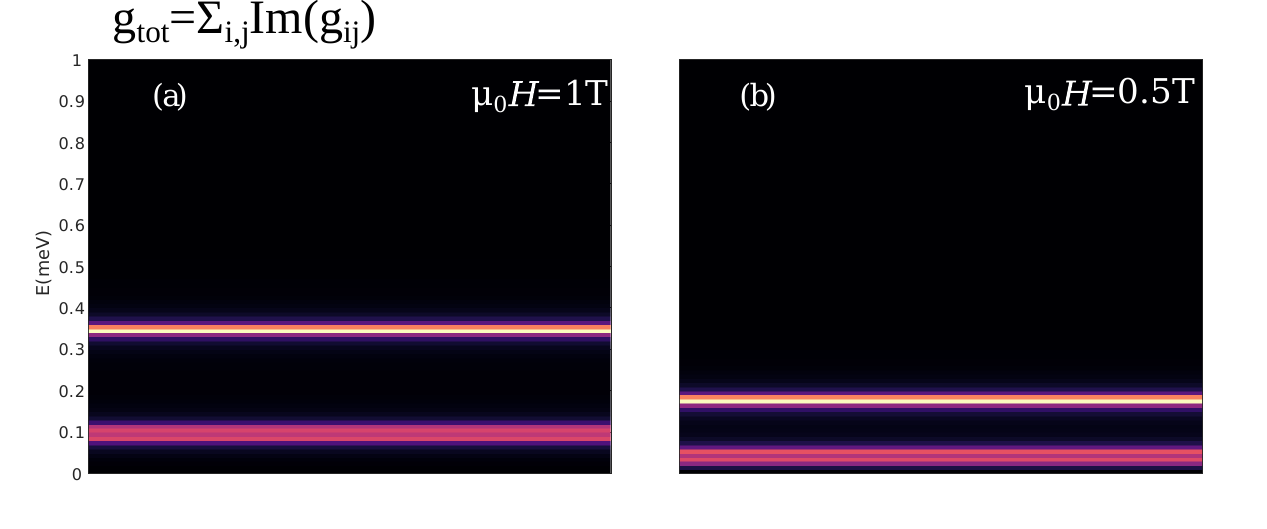}
	\caption{Plots of single-ion response for the two ion excitations at (a)$\mu_0\mathit{H}$ =  1T, (b) 0.5T.}
	\label{figure_5}
\end{figure}

\section{Experiment}

\renewcommand{\thefigure}{S2}
\begin{figure}[t]
	\includegraphics[width=60mm,trim=1cm 1cm 18cm 1cm,clip=true]{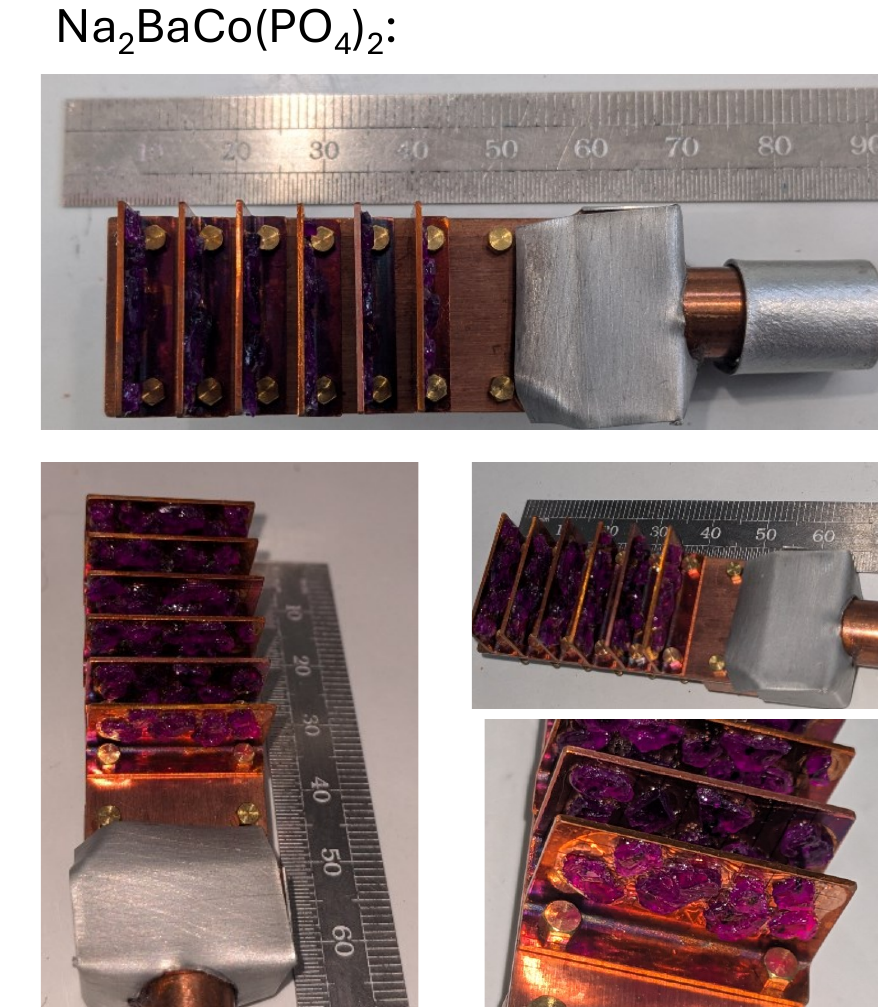}
	\caption{ The coaligned sample mount used in the experiments.  The purple NBCP samples were mounted on copper plates which is the origin of the intense phonons obsever below $\sim$ 30 meV in the high-energy experiments investigating $\mathcal{H}_{SI}$.}
	\label{figure_sample_mount}
\end{figure}

\renewcommand{\thefigure}{S3}
\begin{figure}[t]
	\includegraphics[width=80mm,trim=0cm 3.5cm 0cm 3cm,clip=true]{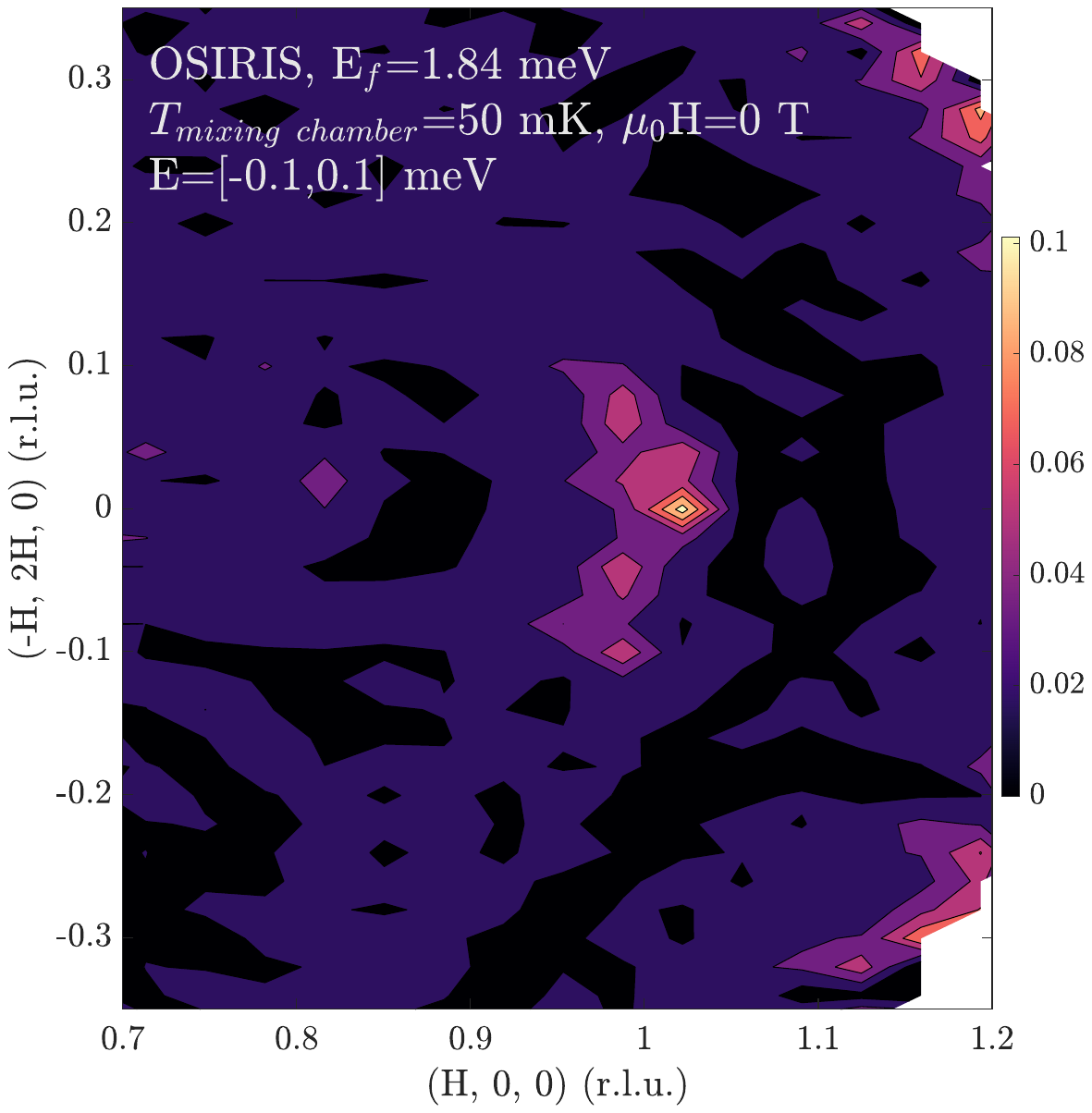}
	\caption{An elastic $E$=0 $\pm 0.1$  meV slice from OSIRIS illustrating the coalignment characterized by the $\vec{Q}$=(100) Bragg peak.}
	\label{figure_elastic_line}
\end{figure}

\textit{Neutron Scattering} - The experiments were performed on the OSIRIS and MAPS spectrometers located ISIS (UK).  OSIRIS is a backscattering spectrometer with E$_{f}$=1.84 meV and was used with an applied magnetic field and dilution refrigerator to probe the low-energy magnetic excitations within the $j_{eff}={1\over 2}$ ground state Kramers doublet.  To fix the single-ion parameters in $\mathcal{H}_{SI}$, we applied the MAPS direct geometry spectromet with E$_{i}$=75 meV with the Fermi chopper spun at 300 Hz.  

\textit{Sample} - The sample consisted of a series of coaligned NBCP samples mounted on copper plates such that reflections of the form (HK0) lay within the horizontal scattering plane.  The sample mount is displayed in Fig. \ref{figure_sample_mount} with an elastic line slice through the OSIRIS data illustrating the (100) Bragg peak shown in Fig. \ref{figure_elastic_line}.  The clustering of intensity around (100) demonstrates the quality of the coalignment.

\textit{Analysis} - We started the paper above by presenting an outline of the data obtained from the 4 T field measurements, comparing it to the theoretical results from our calculations, as illustrated in Figure 2 of the main text. We discuss data processing and detector masking given electronic difficulties that occured during the experiment.  For the magnetic field data sets, several detectors were found to give significantly higher detector counts when compared to nearest neighbor detectors.  We therefore imposed a mask of detectors 4 (for $\mu_{0}H= 1$ T), 13 (for $\mu_{0}H= 4$ T) and 24 (all fields), the effect of which is most visible in Figure 2 [$(a)$ $(b)$] and Figure 4 [$(c)$ $(e)$] of the main text. The data from detector 24 was  removed in all  measurements to account for anomalous readings in the high momentum range. 

\begin{table}[h]
	\begin{ruledtabular} \label{Table_parameter}
		\begin{tabular}{>{\centering\arraybackslash}p{0.25\linewidth} >{\centering\arraybackslash}p{0.25\linewidth} >{\centering\arraybackslash}p{0.25\linewidth} >{\centering\arraybackslash}p{0.25\linewidth}}
			\multicolumn{4}{c}{Field-Independent Parameters} \\ 
			\hline
			Parameter & Value (meV) & Parameter & Value  \\ 
			\hline 
			$\mathcal{J}_1$ & $0.0095$ $\pm$ 0.0005  & $\lambda$ & $-18$ $\pm$ 2 meV \\ 
			$\mathcal{J}_2$ & $0$ & $\alpha$ & $-3/2$  \\ 
			$\mathcal{J}_3$ & $0$ & $\Gamma$ & $-20$  $\pm$ 2 meV \\ 
			\hline 
			\multicolumn{4}{c}{Field-Dependent Parameters} \\ 
			\hline
			$B (T)$   & $h_{\text{MF}}$ (meV) & $\xi$ (\AA$^{2}$/meV) & \\ 
			\hline 
            0 & -0.060 $\pm$ 0.015 & 0.13 $\pm$ 0.010 & \\
            1 & -0.030 $\pm$ 0.015 & 0.090 $\pm$ 0.010 & \\ 
            2 & -0.030 $\pm$ 0.007  & 0.075 $\pm$ 0.010 & \\ 
            3 & 0.040 $\pm$ 0.005 & 0.065 $\pm$ 0.010 & \\ 
			4 & 0.114 $\pm$ 0.003 & 0.06 $\pm$  0.015 & \\ 
		\end{tabular}
	\end{ruledtabular}
	\caption{\label{Table:parameter} Summary of field-independent and field-dependent parameters.}
\end{table}

\renewcommand{\thefigure}{S4}
\begin{figure}[h]
	\includegraphics[ width=1\linewidth]{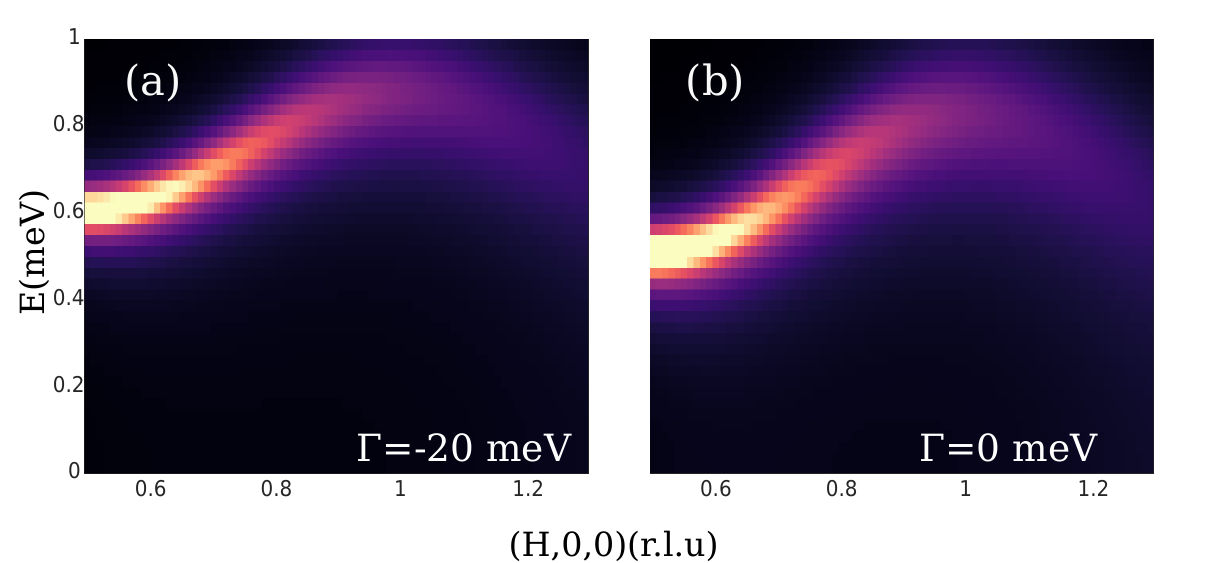}
	\caption{Comparison between the RPA calculation (a)including the distortion parameter and (b) excluding it.}
	\label{figure_6}
\end{figure}

The parameters employed in the RPA calculation for various fields are provided in Table \ref{Table:parameter}. The qualitative interpretation of these parameters may be understood in the context of their impact on the excitation observed. The nearest exchange parameter $\mathcal{J}_1$ may be understood as varying the amplitude of the excitation, while the molecular field $h_{\text{MF}}$ shifts the minimum energy of the excitation given that it is a Zeeman term. Constant momentum cuts for the model with these parameters are plotted on top of the neutron scattering data and are presented in Fig. \ref{figure_slice}. Of notable importance is the distortion parameter $\Gamma$, in whose absence the RPA calculation suggested that a smaller value for $\mathcal{J}_1$ would have been required (see Fig. \ref{figure_6}).  To evaluate the parameter uncertainties in the model we systematically varied each parameter and observed how these affected the model's output in comparison to the data.  Based on this we established a range of parameters where the data and model were still consistent.  Overall, we determined that the parameters could be determined within 5-10 \% error.

\renewcommand{\thefigure}{S5}
\begin{figure}[t]
	\includegraphics[width=1\linewidth,trim=0cm 0cm 0cm 1cm,clip=true]{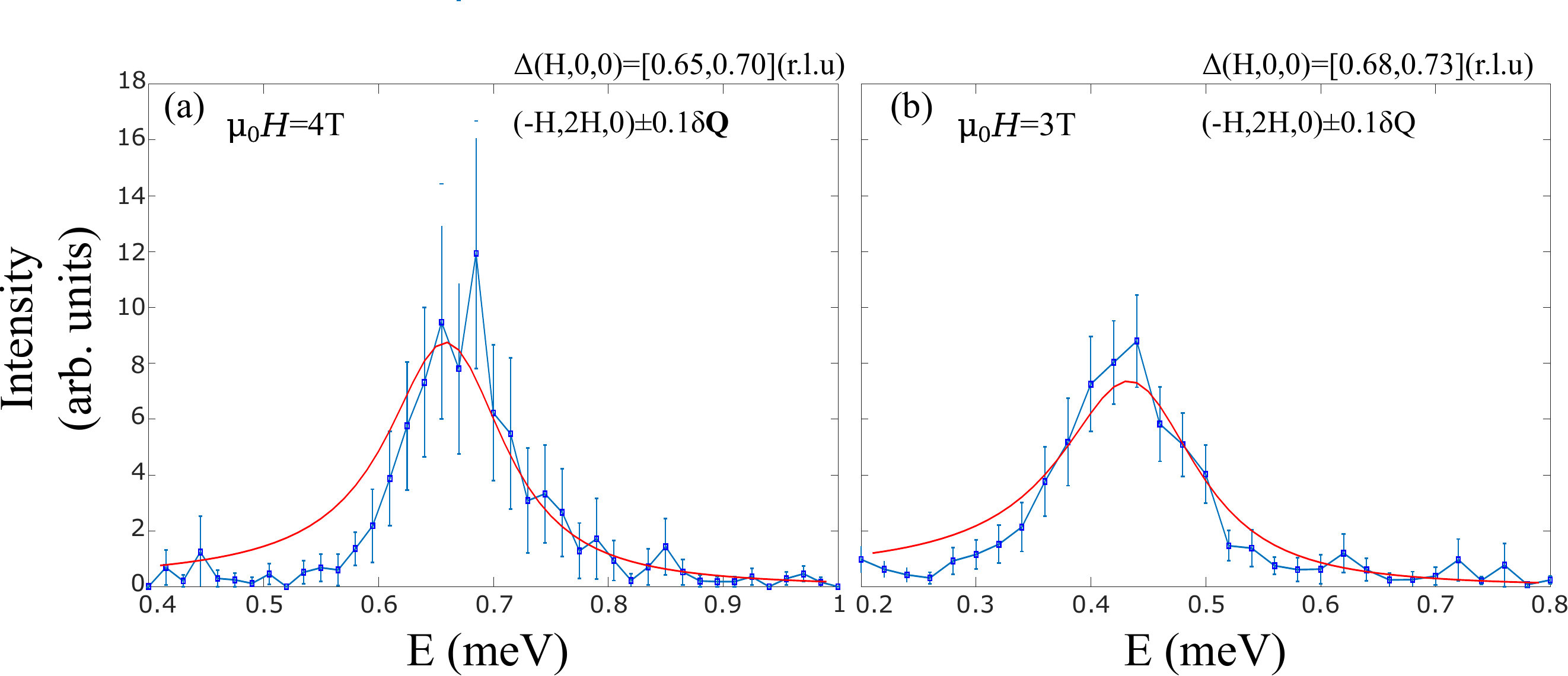}
	\caption{Constant momentum slices of the neutron scattering data (blue) with the overplotted RPA calculation (red) for magnetic fields (a) $\mu_0\mathit{H}$ = 4.0 T, (b) 3.0 T .}
	\label{figure_slice}
\end{figure}

\renewcommand{\thefigure}{S6}
\begin{figure}[t]
	\includegraphics[width=50mm,trim=5cm 5cm 6cm 5cm,clip=true]{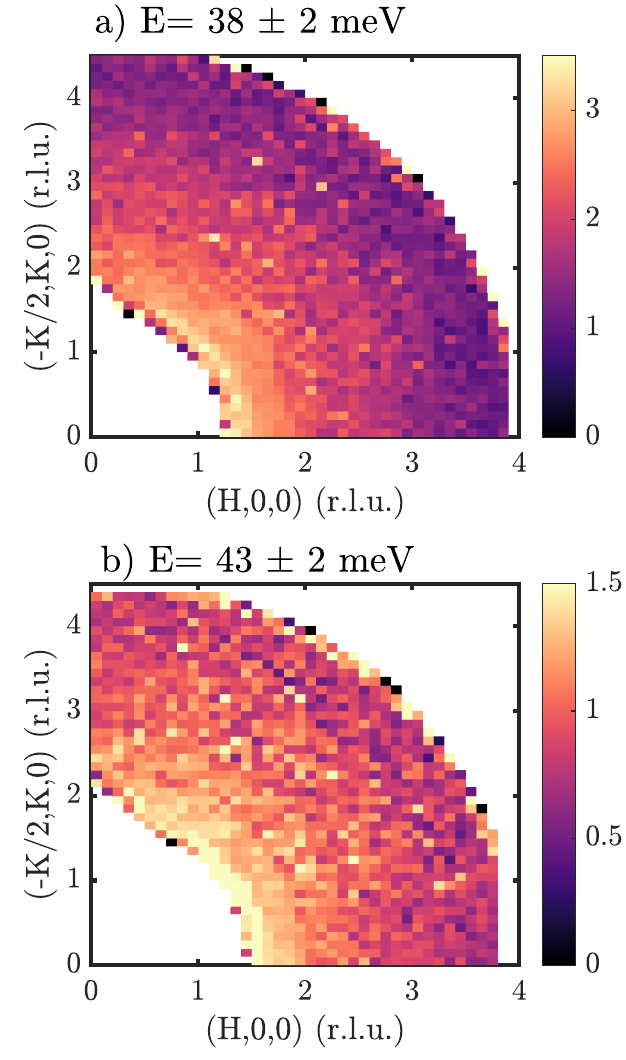}
	\caption{Constant energy slices through the two dispersionless excitations near $\sim$ 40 meV reported in the main text in Figure 1 $(c)$.  Decaying intensity with increasing momentum transfer can be observed and also the isotropic nature of the scattering expected from single-ion like excitations.}
	\label{figure_constant_energy}
\end{figure}

\renewcommand{\thefigure}{S7}
\begin{figure}[t]
	\includegraphics[width=60mm,trim=4cm 7cm 5cm 7cm,clip=true]{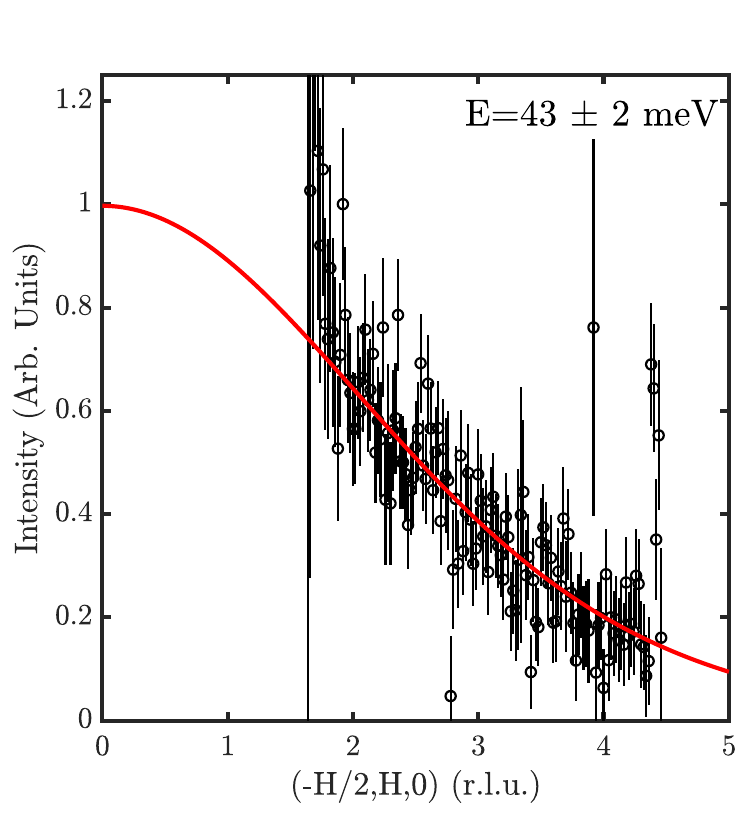}
	\caption{The isotropic magnetic form factor from Co$^{2+}$ is over-plotted the excitation at 43 meV confirming the magnetic nature.}
	\label{figure_form_factor}
\end{figure}

The spin-orbit coupling $\lambda$ and distortion $\Gamma$ parameters in $\mathcal{H}_{SI}$ were derived from the high-energy MAPS data illustrated in Fig. 1 $(c)$ of the main text.  Kramers doublets cannot be split by a time reversal symmetry preserving crystalline electric field, and therefore investigating spin-orbit crystal field excitations are required to derive $\lambda$ and $\Gamma$.  As noted in the main text, two distinct dispersionless excitations were observed near $\sim$ 40 meV.  These excitations are magnetic and decay with momentum transfer.  This is illustrated in Fig. \ref{figure_constant_energy} which plots constant energy slices through the two dispersionless excitations near $\sim$ 40 meV.  The scattering intensity decays with increasing momentum transfer and is also isotropic as expected from single-ion crystal type excitations.  In Fig. \ref{figure_form_factor} we plot the momentum dependence of the intensity of the 43 meV excitation.  Overplotted in red is the isotropic Co$^{2+}$ form factor.


%